\documentclass[twocolumn,twocolappendix]{aastex63}

\received{June 1, 2019}
\revised{January 10, 2019}
\accepted{\today}
\submitjournal{ApJ}

\def\be{\begin{equation}}
\def\ee{\end{equation}}
\def\CF3{{\sc cosmicflows-3}}
\usepackage{rotating}
\usepackage{multirow}
\usepackage{CJKutf8}
\usepackage{graphicx}	
\usepackage{amsmath}	
\usepackage{amssymb}	

\usepackage{newtxtext,newtxmath}

\usepackage[T1]{fontenc}
\usepackage{ae,aecompl}

\usepackage{color}

\shorttitle{Cosmic flow and mocks of CF4TF}
\shortauthors{Qin et al.}
\graphicspath{{./}{figures/}}

\begin{document}

\title{Cosmic Flow Measurement and Mock Sampling Algorithm of Cosmicflows-4 Tully-Fisher Catalogue}

\correspondingauthor{Fei Qin}
\email{feiqin@kasi.re.kr}

\author{Fei Qin}
\altaffiliation{Korea Astronomy and Space Science Institute, Yuseong-gu, Daedeok-daero 776, Daejeon 34055, Republic of Korea.}

\author{David Parkinson}
\affiliation{Korea Astronomy and Space Science Institute, Yuseong-gu, Daedeok-daero 776, Daejeon 34055, Republic of Korea.}

\author{Cullan Howlett}
\affiliation{School of Mathematics and Physics, The University of Queensland, Brisbane, QLD 4072, Australia.}

\author{Khaled Said}
\affiliation{School of Mathematics and Physics, The University of Queensland, Brisbane, QLD 4072, Australia.}



\begin{abstract}

Measurements of cosmic flows enable us to test whether cosmological models can accurately describe the evolution of the density field in the nearby Universe. In this paper, we measure the low-order kinematic moments of the cosmic flow field, namely bulk flow and shear moments,  using the Cosmicflows-4 Tully-Fisher catalogue (CF4TF). To make accurate cosmological inferences with the CF4TF sample, it is important to make realistic mock catalogues. We present the mock sampling algorithm of CF4TF. These mock can accurately realize the survey geometry and luminosity selection function, 
     enabling researchers to explore how  these systematics  affect the measurements. These mocks can also be further used to 
     estimate the covariance matrix and errors of power spectrum and two-point correlation function in future work. In this paper, we use the  mocks to test the cosmic flow estimator and find that the measurements are unbiased. 
        The measured bulk flow in the local Universe is 376 $\pm$ 23 (error) $\pm$ 183 (cosmic variance) km s$^{-1}$ at depth $d_{\text{MLE}}=35$ Mpc $h^{-1}$, to the Galactic direction of $(l,b)=(298\pm 3^{\circ},  -6\pm 3^{\circ})$. Both the measured bulk and shear moments are consistent with the  concordance $\Lambda$ Cold Dark Matter cosmological model predictions.

\end{abstract}



\section{Introduction} \label{sec:intro}

Observations of galaxies indicate that there are  fluctuations in 
the density field of local Universe (\citealt{Jarrett2004} and references therein). The gravitational effects of the density   perturbations  exert additional velocity components to the galaxies' Hubble recessional velocities, called `peculiar velocities'. These peculiar velocities are good probes  of the density field, enabling us to constrain cosmological parameters and test cosmological models.

Due to the peculiar motions of galaxies, the   apparent (or inferred) distance of a galaxy, $d_z$ is different from its true comoving distance, $d_h$. This difference is measurable, and quantified by the `logarithmic distance ratio' for that galaxy,  defined as
\be\label{logd}
\eta \equiv \log_{10}\frac{d_z}{d_h}~.
\ee
The $\eta$ of late-type galaxies can be measured from the Tully-Fisher relation \citep{Tully1977,Strauss1995,Masters2008,Hong2014}, while for early-type galaxies, $\eta$ can be measured from the Fundamental Plane (\citealt{Djorgovski1987,Dressler1987,Strauss1995,Magoulas2012,Springob2014,Said2020}; Howlett et al. 2021 in prep). The peculiar velocity of a galaxy can be estimated from its log-distance ratio using velocity estimators \citep{Davis2014,Watkins2015,Adams2017}.

The cosmic flow field in the local Universe arises from the peculiar velocities of galaxies. Measuring the low-order kinematic moments of the cosmic flow field, i.e. bulk flow and shear moments, from  peculiar velocity surveys and comparing them to the $\Lambda$ Cold Dark Matter theory predictions enables us to test whether the  theory accurately describe the motion of galaxies on large scales. 
In previous work, the bulk and shear moments are commonly measured using two  different methods: maximum likelihood estimation (MLE, \citealt{Kaiser1988}); minimum variance (MV) estimation \citep{Watkins2009, Feldman2010}. The measurements generally agree so far with the  
$\Lambda$ cold dark matter model ($\Lambda$CDM)  prediction
\citep{Kaiser1988,Lister1989,Jaffe1995,Nusser1995,Parnovsky2001,Nusser2011,Turnbull2012,Ma2013,Ma2014,Hong2014,Scrimgeour2016,Qin2018,Qin2019a,Boruah2020,Qin2021a,Stahl2021}.

In this paper we will measure the bulk flow and shear moments using the peculiar velocity catalogue,  Cosmicflows-4 Tully-Fisher catalogue (CF4TF, \citealt{Kourkchi2020}), and compare the measurements to the $\Lambda$CDM predictions to test the model.
CF4TF is the currently largest full sky catalogue of Tully-Fisher galaxies, enables us to more accurately measure the bulk and shear moments and  
the measurements are less affected by the an-isotropic sky coverage, comparing to previous work.

In addition, as CF4TF is one of the key components of the future full Cosmicflows-4 catalogue, we present  the  mock sampling algorithm for CF4TF in this paper. Combining these CF4TF mocks with the 6dFGSv mocks \citep{Qin2018,Qin2019b}, we can obtain mock catalogues 
for the two largest 
subsets of the final full Cosmicflows-4 catalogue.   
Our mocks can model the luminosity selection, survey geometry of the real surveys, enabling researchers  to explore how  these systematics  affect the measurements (\citealt{Qin2018,Qin2019a}; Howlett et al. 2021 in prep). In future work, these mocks  can be used to study the power spectrum and two-point correlation of velocities of Cosmicflows-4 catalogue. They are the key to estimate the covariance matrix and errors of these measurements. They can also be used to test the methods of these measurements to identify any possible biases.  The mock catalogues underlying this article will be shared on a reasonable request to the corresponding author.   In this paper, these mock are used to test the cosmic flow (bulk and shear moments) estimator to explore how well the estimator recover the true moments.

The  paper  is  structured  as  follows: in Section \ref{sec:data}, we introduce the CF4TF data. In Section \ref{sec:mock}, we introduce the L-PICOLA simulation \citep{Howlett2015a,Howlett2015bs}  and present the mock sampling algorithm. In Section \ref{sec:BKm}, we introduce the cosmic flow estimator and test the estimator using mocks. We  present  the  final  results and discussion in  Section \ref{sec:Result}. A conclusion is presented in Section \ref{sec:conc}.

We adopt a spatially flat $\Lambda$CDM cosmology as the fiducial model. The cosmological parameters  are: $\Omega_m=0.3121$,  $\Omega_{\Lambda}=0.6825$ and $H_{0} = 100$ $h$ km s$^{-1}$ Mpc$^{-1}$ with $h=0.6751$.

\section{Data} \label{sec:data}

\begin{figure*} 
 \includegraphics[width=175mm]{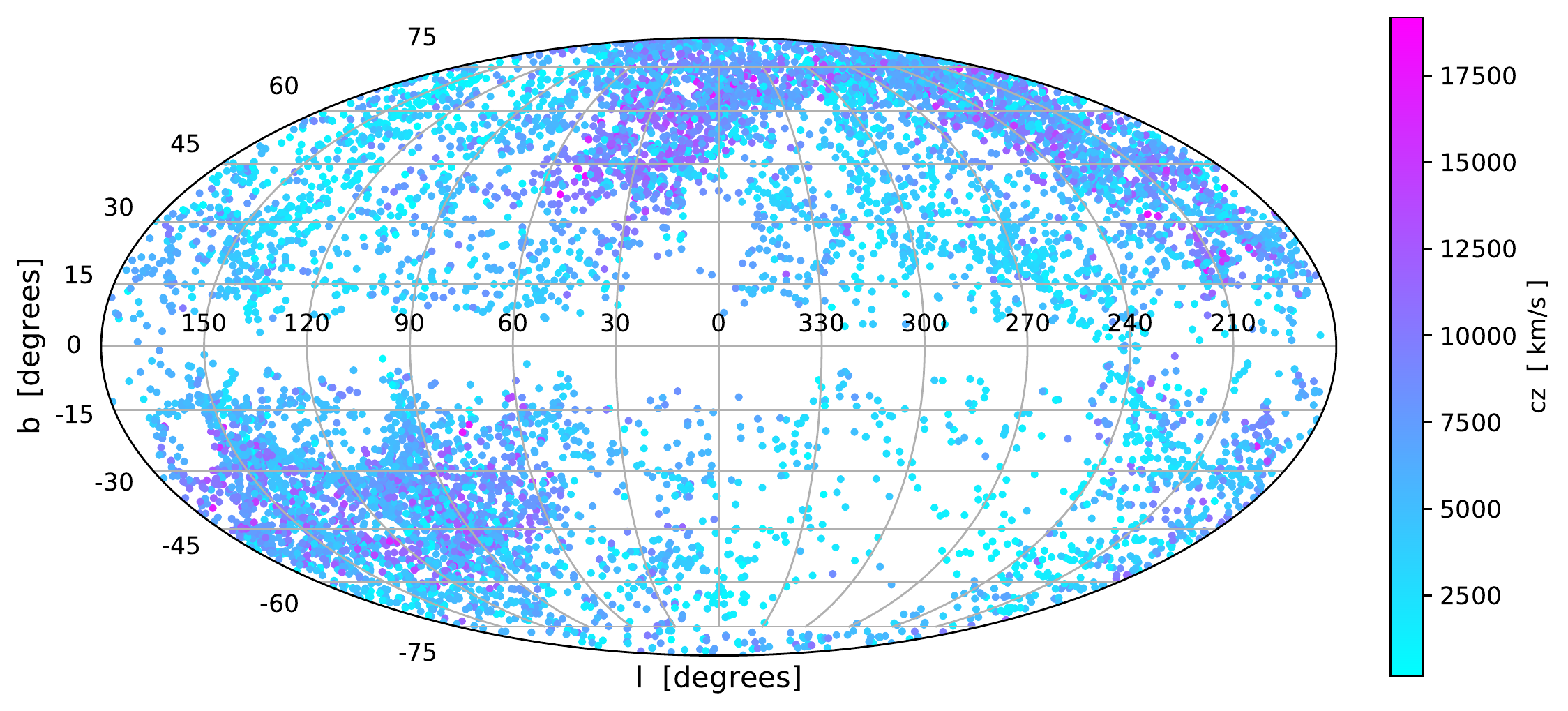}
 \caption{The sky coverage of the CF4TF galaxies under Galactic coordinates. The colors of the dots represent the redshift of the galaxies, according to the color bar. }
 \label{lb}
\end{figure*}

\begin{figure} 
\centering
 \includegraphics[width=\columnwidth]{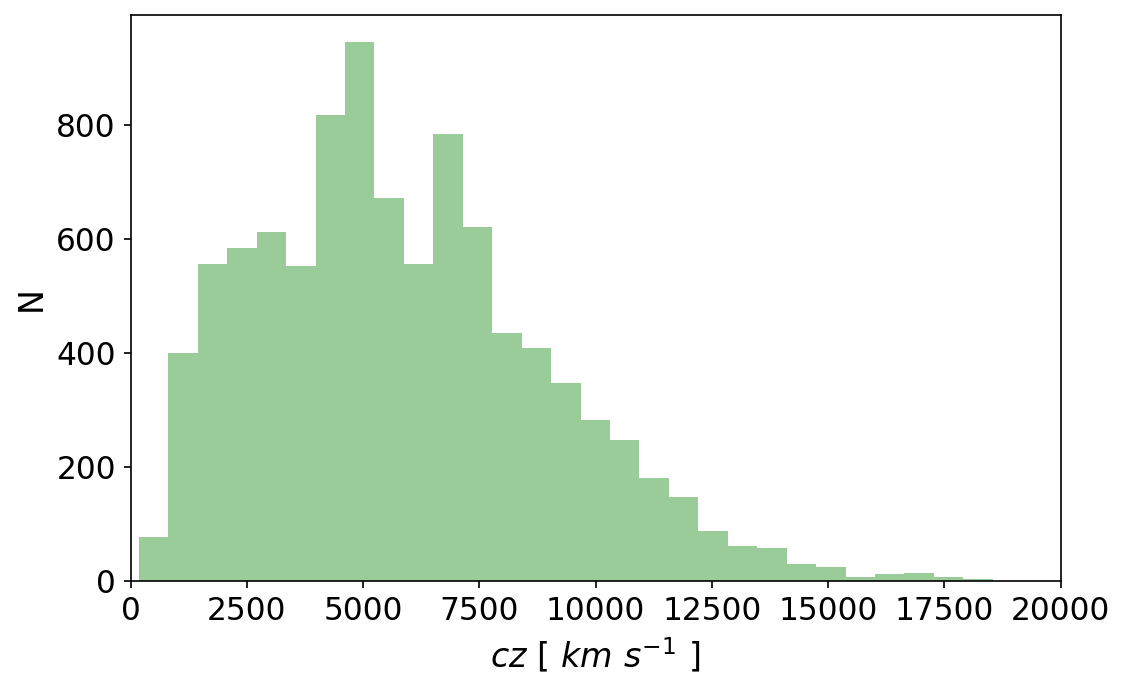}
 \caption{ The redshift distribution  of the CF4TF galaxies. }
 \label{histcz}
\end{figure}

The Cosmicflows-4 Tully-Fisher catalogue (CF4TF, \citealt{Kourkchi2020}) is a full-sky catalogue of 9790 galaxies. 
The sky coverage of the CF4TF galaxies under Galactic corrdinates is shown in Fig.\ref{lb}. 
The redshift of the CF4TF galaxies reaches ~20000 km s$^{-1}$, the redshift distribution is shown in Fig.\ref{histcz}.  
The distances of galaxies are measured using the Tully-Fisher relation \citep{Tully1977,Kourkchi2020}, which is a linear relation between the H\thinspace{\protect\scriptsize I} rotation widths and photometry magnitudes. 

The H\thinspace{\protect\scriptsize I} data is taken from the following four catalogues. 
The primary catalogue is the All Digital H\thinspace{\protect\scriptsize I}  catalogue (ADHI, \citealt{Courtois2009} ), the galaxies with H\thinspace{\protect\scriptsize I} line widths uncertainties $\le 20$ km s$^{-1}$ are selected. In the ADHI, the  galaxies below declination $\delta=-45^{\circ}$ are observed by the Parkes
Telescope \citep{Courtois2011}, the number density of galaxies is lower in this region. 
Most  of  the  remainder is the Arecibo Legacy Fast ALFA Survey (ALFALFA, \citealt{Haynes2018,Haynes2011}), mainly distributed in declination range of $[0^{\circ},~38^{\circ}]$ and the galaxies with H\thinspace{\protect\scriptsize I}  spectrum signal-to-noise ratio SNR>10 are selected.
The rest are the
Springob/Cornell catalog \citep{Springob2005} 
and the  Pre-Digital H\thinspace{\protect\scriptsize I} catalog \citep{Fisher1981,Huchtmeier1989}.

The u, g, r, i and z bands photometry is taken from the Sloan Digital Sky Survey (SDSS) Data Release 12 (DR12, \citealt{York2000}). The $w_1$ and $w_2$ bands data is taken
from the Wide-field Infrared Survey Explorer (WISE, \citealt{Wright2010}). The i-band absolute magnitude cut $M_i\le -17$ Mag is applied to the data \citep{Kourkchi2020,Kourkchi2020a}.

\section{Mocks} \label{sec:mock}

\subsection{L-PICOLA simulation}

Producing  realistic mock surveys that
model non-linear gravitational interactions requires N-Body
simulations. 
However, such full N-body simulations, which
model the gravitational forces and motions down to a fine resolution, can require an extensive HPC infrastructure and a large
amount of CPU-time. To make a large-ensemble of mock surveys quickly and efficiently, we use the COmoving Lagrangian
Acceleration (COLA) approach \citep{Tassev2013}    implemented in the \textsc{l-picola} code \citep{Howlett2015a,Howlett2015bs}. 

 We use \textsc{l-picola} to generate 250 dark matter simulations. Though approximate, this method and code have been demonstrated to reproduce the clustering of dark matter particles and halos well on all scales of interest for galaxy and peculiar velocity clustering \citep{Howlett2015a,Howlett2015bs,Blot2019}, at $\gtrsim 95\%$ accuracy for both the power spectrum and bispectrum at $k=0.3h\mathrm{Mpc^{-1}}$. Halos are identified in these simulations using the 3D Friends-of-friends algorithm in the VELOCIraptor code \citep{Elahi2009}. This code has also been demonstrated to recover realistic halos, subhalos and halo merger trees (although we only the the first of these here), in agreement with a variety of other halo-finders \citep{Onions2012,Knebe2013}.

 The fiducial cosmology is given by $\Omega_m=0.3121$, $\Omega_b=0.0488$, $\sigma_8=0.815$, and $h=0.6751$.
The redshift of the simulations is at $z=0$.  The boxsize is 1800 Mpc h$^{-1}$. The number of particles in each simulation is 2560$^3$. The minimum halo mass is $\sim$5$\times$10$^{11}M_{\sun}h^{-1}$ (20 particles per halo).

\subsection{Mock sampling algorithm for CF4TF}

We have 250 simulation boxes in total, 
each simulation provides us with 
the mass $m_{hl}$ of halos,
 Cartesian positions $[x,y,z]$ and velocities $[v_x,v_y,v_z]$ of halos. We divide each simulation box into eight  identical sized cubes. In each  cube, the observer is placed at a galaxy close to the center of that cube. Therefore, we have generated 2000 mock CF4TF catalogues in total.

The halos in the simulations are parent halos, it is required to generate subhalos and galaxies from these parent halos.  
The starting point is the halo concentration, defined as
\be 
c_v\equiv\frac{r_v}{r_s}
\ee 
which   characterises  the mass distribution in a halo. $r_v$ is the virial  radius  of  a  halo.  $r_s$ is  the  break  radius  between  an  inner and outer density  profile in a halo. 
  $c_{v}$ is computed purely as a function of parent halo mass $m_{v}$ using the fitted relationship from \cite{Prada2012}, calibrated on high resolution N-body simulations. The virial radius is also computed using the parent halo mass and assuming the halo is both spherical and has a density $200 \times$ the critical density of the Universe.
 Given  $c_v$ and $r_v$, one can generate the subhalos and galaxies using the Navarro–Frenk–White (NFW) profile \citep{Navarro1997}. 

 The subhalo generation algorithm 
is based on \citealt{Vale2004, Conroy2006,Howlett2015,Howlett2017,Qin2019b} and detailed below:

1. First, we assume a power law distribution for the mass ratio between a subhalo and its parent halo. From this power law distribution, we can compute the expected number of subhalos as 
\be  \label{nsubs}
\lambda_{Poi}=\int^1_{f_{min}} Af^{-\alpha_h}_M df_M
\ee 
where $f_M$ is the mass ratio between
the subhalo and its parent.   $f_{min}$ is the minimum mass ratio we consider, which we set, based on the minimum halo mass of the simulation (see Section 3.1), to 20 times the dark matter particle mass divided by the parent halo mass (i.e., a fixed minimum mass, such that the minimum mass \textit{ratio} varies based on the parent halo). 
The free parameters $A$ and $\alpha_h$ will be fitted by matching the density power spectrum of mocks to real data.

2. To add realism to the sampling process, the actual number of subhalos in each parent halo $N_{sub}$ is generated assuming a Poisson distribution with mean $\lambda_{Poi}$. The mass ratios of the $N_{sub}$ subhalos are generated by drawing uniform random numbers in the interval $[0,~1]$, and inverting the quantile distribution for Eq.~\ref{nsubs}. This results in mass ratios in the interval $[f_{min},~1]$. The subhalo's masses are then obtained by multiplying the set of $f_M$ by the mass of the parent halo. As the largest value of $f_{M}$ generated for the $N_{sub}$ subhalos is 1, the masses of the subhalos will not be larger than their parent halos. 


3. The positions of subhalos are generated from the NFW profile
following the arguments in   \citealt{Robotham2018},   and rely on the concentration and virial radius for each parent halo computed as described previously.  We generate $N_{sub}$ random numbers in the interval $R\in[0,1]$, then calculate the position parameter \citep{Robotham2018}
\be 
p=R \left(  \ln(1  + c_v )-\frac{c_v} {1  + c_v }  \right)
\ee 
and the radius
\be 
r=-\frac{r_v}{c_v}\left(  1+ \frac{1}{W_0(-e^{-p-1})}     \right)
\ee 
which gives the radial position of the subhalos relative to the   center of their parent halo. Here  $W_0$ is the Lamber Function\footnote{We use PYTHON function scipy.special.lambertw.}.
Generating $N_{sub}$ random numbers in the intervals $[-\pi,~\pi]$ and  $[0,~2\pi]$ as the polar coordinates of the subhalos, respectively. Then we convert $r$ into Cartesian coordinates using the polar coordinates. Finally we add them to the  Cartesian positions of their parent halo and subtract  the observer's position to obtain the final positions of subhalos.

4. The velocities of subhalos are calculated from \citep{Navarro1997,Howlett2015}
\be 
v=\sqrt{\frac{Gm_{v}s}{r_v}}
\ee 
where $G$ is the Newton Gravitational Constant and $s$ is the  ratio between the velocities of subhalos and circular velocity of their parent halo, given by \citep{Navarro1997,Howlett2015}
\be 
s = \frac{1}{q}\frac{ \ln(1 + c_vq) - \frac{c_vq}{1 + c_vq}        }
{
\ln(1  + c_v )- \frac{c_v}{1+c_v}
}
\ee 
Then we randomly draw the Cartesian components of the velocities $v_x$, $v_y$ and $v_z$ for the subhalos from the Gaussian function with mean equals to the velocity  of their parent halo  and standard deviation (std) equals to $v/\sqrt{3}$ \citep{Howlett2015}.

5. The i-band absolute magnitude  is generated using the luminosity function (see Appendix \ref{sec:lumi} for more discussion about the i-band luminosity function). We re-order these magnitudes in descending order, then assign them to the halos (both parent and subhalos) in descending order of  mass.  We add additional scatter in this one-to-one assignment via a standard deviation parameter $\sigma_{logM}$ to account for the expected scatter between halo/sub-halo mass and galaxy luminosity. This is used to draw a ``proxy" luminosity for the pairwise-matching of halo mass and luminosity using a Gaussian centred on the actual simulated luminosity. Note that this ``proxy'' is used only for the matching, the simulated luminosity is the one actually stored and used for later applications of the mocks.  Then, we obtain mock galaxies. The normalization factor $\phi_{\star}$ of luminosity function will be fitted by matching the density power spectrum of mocks to real data. 

6. We calculate the sky completeness of the CF4TF survey by splitting into 5 distinct patches (see, Appendix \ref{sec:patch}) and comparing to the 2M++ redshift survey \citep{Lavaux2010}. To do this, the CF4TF and 2M++ are gridded using HEALPix \citep{healpy1,healpy2}. For each HEALPix pixel, we then calculate the ratio of galaxies in CF4TF and 2M++, treating this as the completeness. However there are two caveats to this. First, 2M++ is not quite uniform, it is 1 $K$-band magnitude deeper in the SDSS and 6dF regions, which we correct by normalising the sum of the values in the HEALPix pixels in these regions to be the same. Second, we would not expect every galaxy with a redshift in 2M++ to be capable of providing a TF distance. However, as we are only interested in measuring the completeness of the CF4TF measurements in one region of the sky relative to other regions, the completeness ratio is then normalized such that the mean completeness of the two patches containing ALFALFA data is one. If the completeness of a pixel is greater than one, it is set to be one. 

We note that this doesn't give an absolute measure of completeness (i.e., given all the galaxies which \textit{could} have TF measurements in a region of the real Universe, how many actually \textit{have} measurements in CF4TF), but it does provide a relative measure, which is all that is needed for the mocks given the following steps. Importantly, this procedure removes any signatures of large-scale structure from the completeness mask, as these structures should be present in both CF4TF and 2M++. 

7. For each mock, the simulated galaxies are sub-sampled to match the smoothed redshift distribution of CF4TF data in each of the five distinct sky patches separately. Given the relative sky completeness above, this downsampling ensures the mocks have comparable numbers of objects to those we actually have TF measurements for, and that the completeness in both the angular and radial coordinates is representative of the real Universe. 

8. In a given redshift bin of the real CF4TF data, we use the Gaussian kernel distribution function\footnote{We use the \textsc{python} package  scipy.stats.gaussian-kde.} (KDE) to smooth the distribution of $\epsilon$ (denotes the measurement error of log-distance ratio $\eta$) of that redshift bin. Then we calculate the spline cumulative distribution function (CDF) from the Gaussian KDE for $\epsilon$ of that redshift bin.
The spline CDF is used to
generate  measurements errors  of log-distance ratios for mock galaxies of that redshift bin.
We repeat this step for all the other redshift bins to obtain  the measurements errors of log-distance ratio $\epsilon_{mock}$ for all mock galaxies.  

9. The measured log-distance ratio  for each
mock galaxy is generated  using a Gaussian function centred on the true log-distance ratio $\eta_t$ with standard deviation $\epsilon_{mock}$.\footnote{  The true log-distance ratio of a mock galaxy is given by
$ 
\eta_t=\log_{10}\frac{D_m}{D_h}
$,
where $D_h$ is calculated from the true position of the mock galaxy. $D_m$ is calculated from its `observed' redshift $z_m$, given by 
$ 
z_m=(1  + z_t)(1  + v_t/c) - 1 
$
where $z_t$ is the true redshift calculated from the true position. $v_t$ is the true line-of-sight velocity of the mock galaxy calculated from its true position and velocity.}  As the measured log-distance ratios are centred on the true values, our mocks do not reproduce possible systematics such as Malmquist bias that may affect the real data. To do so would require instead simulating the observed quantities of the Tully-Fisher relationship, and running these through the same pipeline/fitting procedure as the real data. However, such a procedure was already performed when obtaining the CF4-TF data in \cite{Kourkchi2020}, so here we assume the Malmquist bias in the data has been corrected for appropriately, and the mocks only need to reproduce the remaining effects of cosmic variance and measurement errors.

\begin{figure} 
\centering
 \includegraphics[width=\columnwidth]{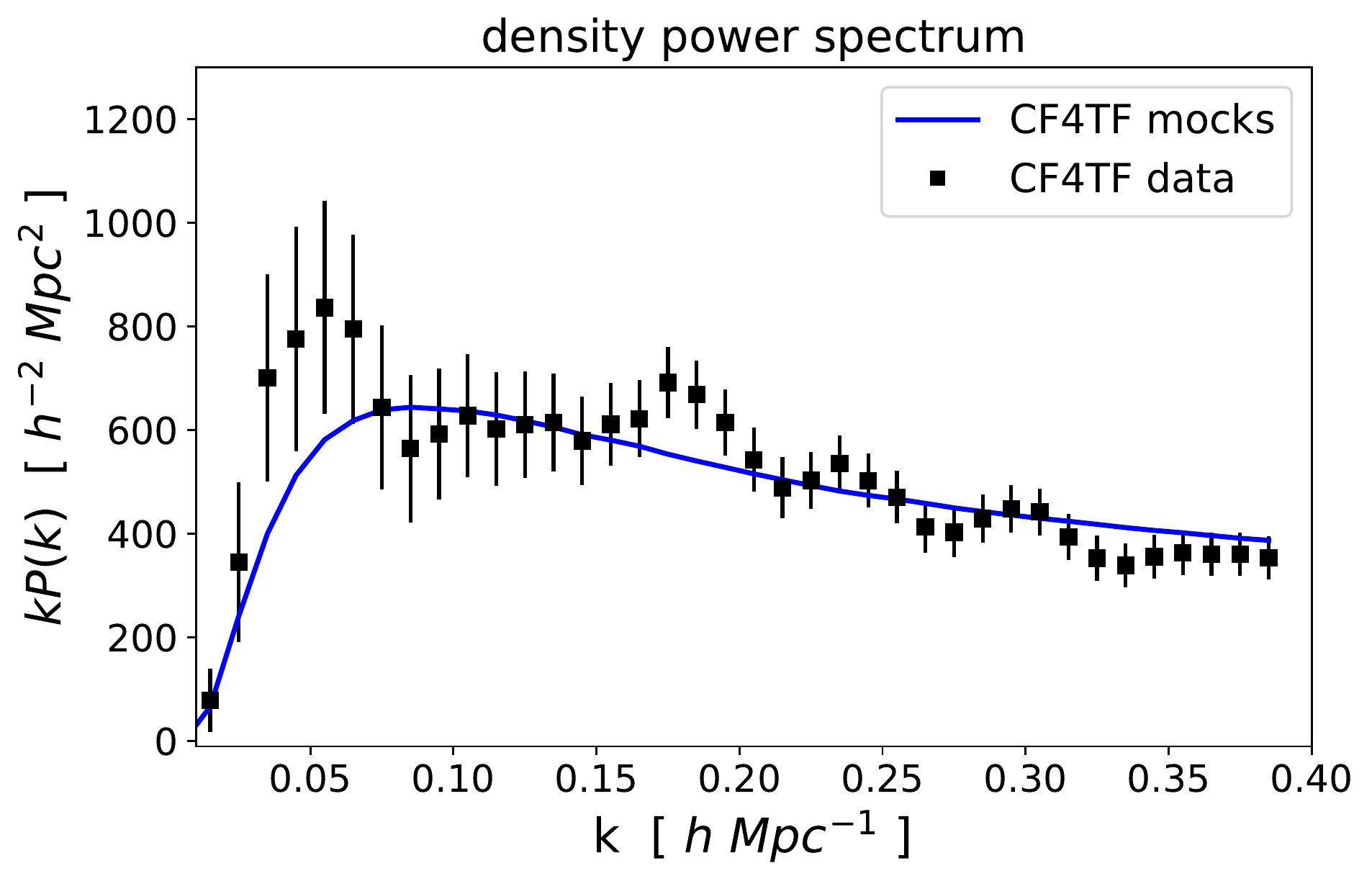}
 \caption{ Comparing the density power spectrum of mocks to data. The blue curve is the average of the density power spectrum of 2000 mocks. The black filled squares are the density power spectrum measured from the real CF4TF data. The $\chi^2$/d.o.f between the data and mock average is  39/34.}
 \label{psmocks}
\end{figure}

10. Finally, we can measure the   redshift-space  density power spectrum of the 2000 mocks using the method\footnote{ In this method, the galaxies are gridded in using their observed redshifts, then using Eq.(2.1.3) of  \citealt{feldman1994} and Eq.6 of \citealt{Bianchi2015} as well as Eq.11 of \citealt{Yamamoto2006} to estimate the density power spectrum. This means that we are mesuring the redshift-space rather than real-space power spectrum, which would need to be accounted for if we were performing any theoretical modelling, but avoids the impact of Malmquist bias or other biases that would arise if one were using the true distance to place the galaxies on the grid.
}   in \citealt{Howlett2019} and \citealt{Qin2019b}. As shown in Fig.\ref{psmocks}, by comparing the average of these measured power spectrum (blue curve) to the density power spectrum of the real CF4TF data (black filled squares) to find the best values for the parameters $A$, $\alpha_h$, $\sigma_{logM}$ and $\phi_{\star}$, as presented in Table \ref{hodpms}. The $\chi^2/d.o.f=39/(38-4)$ indicates that  the  mocks  are  in  excellent  agreement  with  the  data.

\begin{table}   \centering
\caption{The best fit values of the parameters used in the mock sampling algorithm.}
\begin{tabular}{|c|c|c|c|}
\hline
\hline
$A$     &   $\alpha_h$   &   $\sigma_{logM}$   &   $\phi_{\star}$\\
\hline

2.882     & 0.102 & 2.121 & 0.005 \\

\hline
\end{tabular}
\label{hodpms}
\end{table}

\begin{figure*} 
 \centering
 \includegraphics[width=80mm]{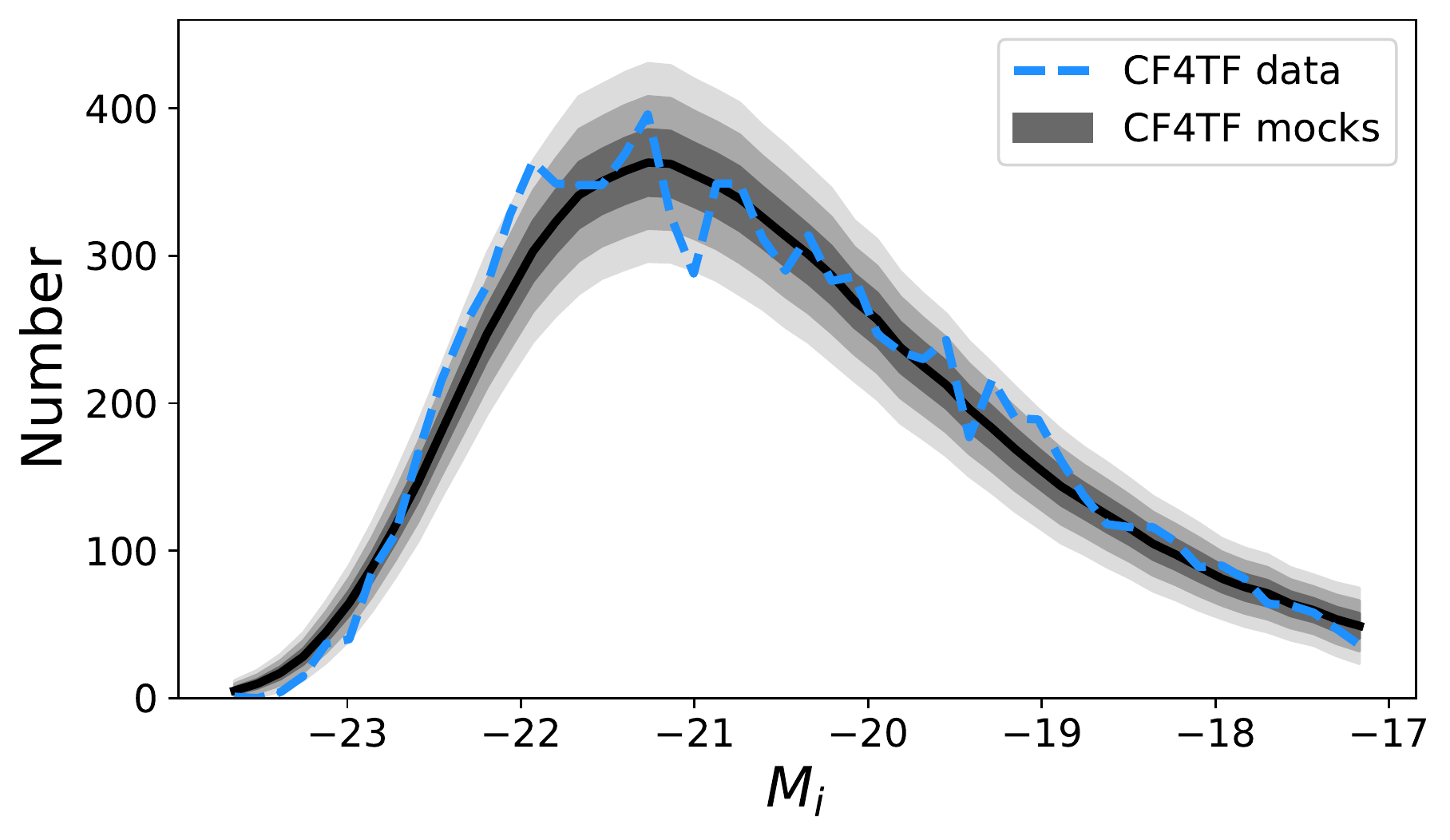}
\includegraphics[width=80mm]{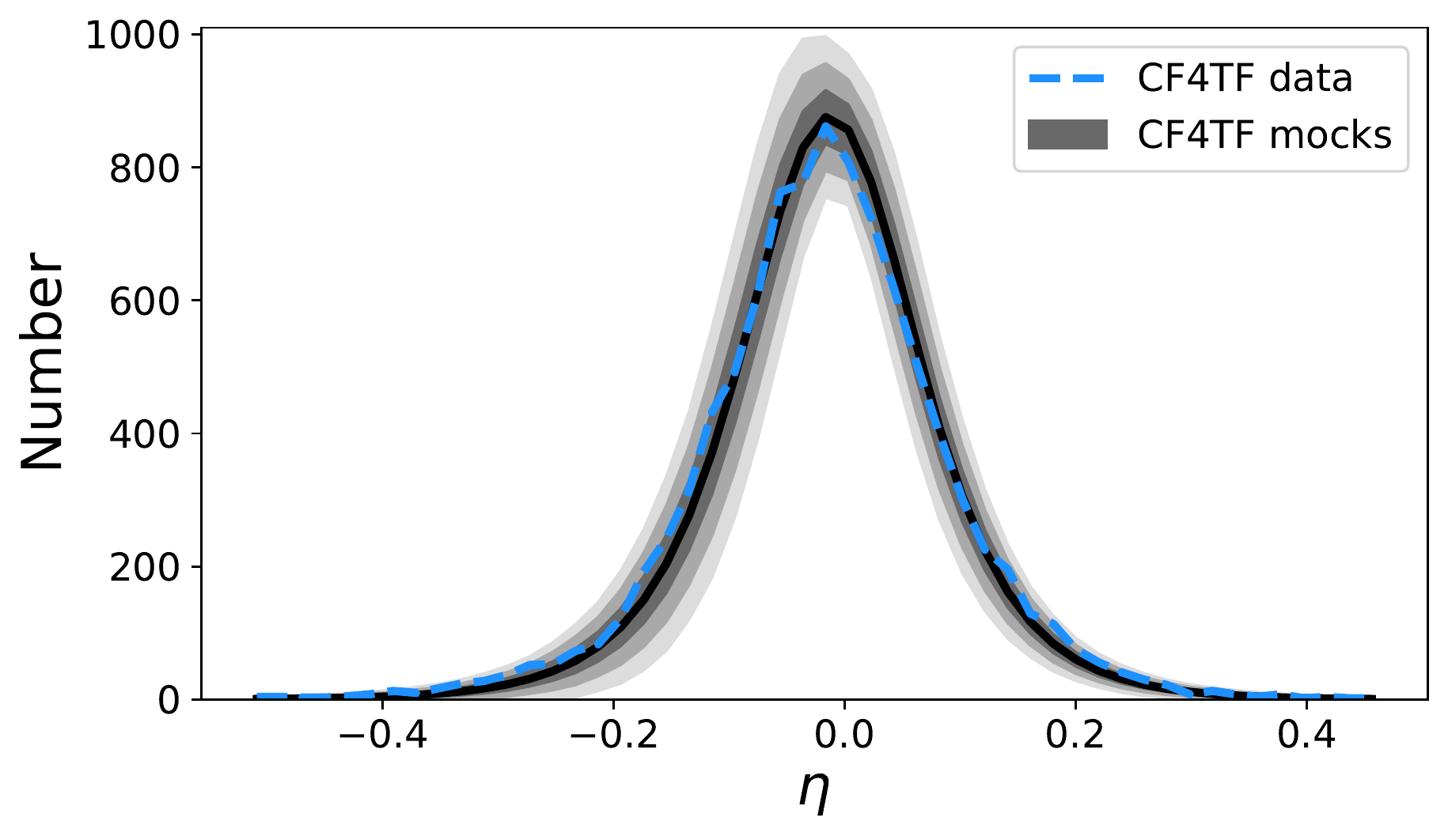}
\includegraphics[width=80mm]{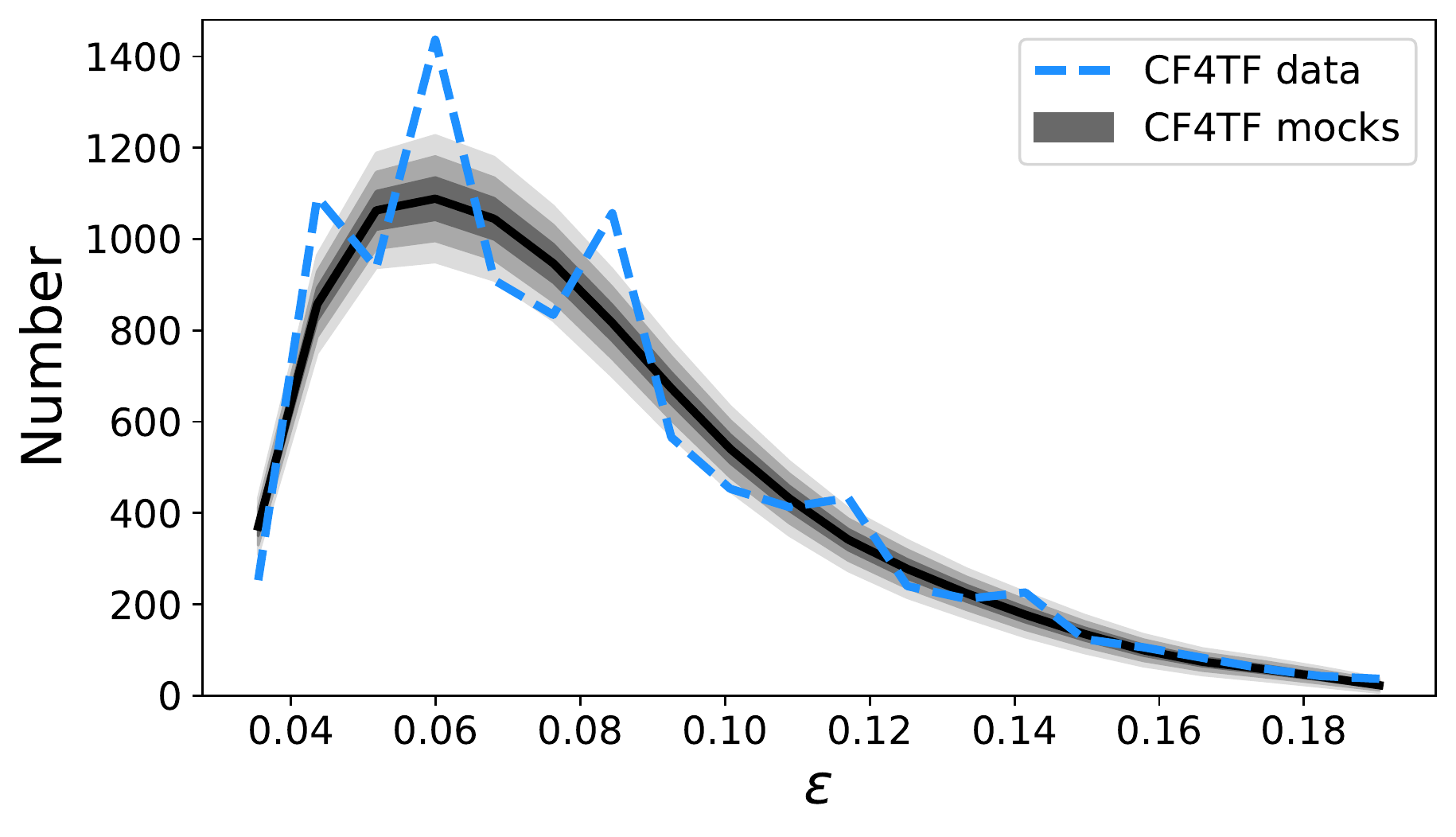}
  \includegraphics[width=80mm]{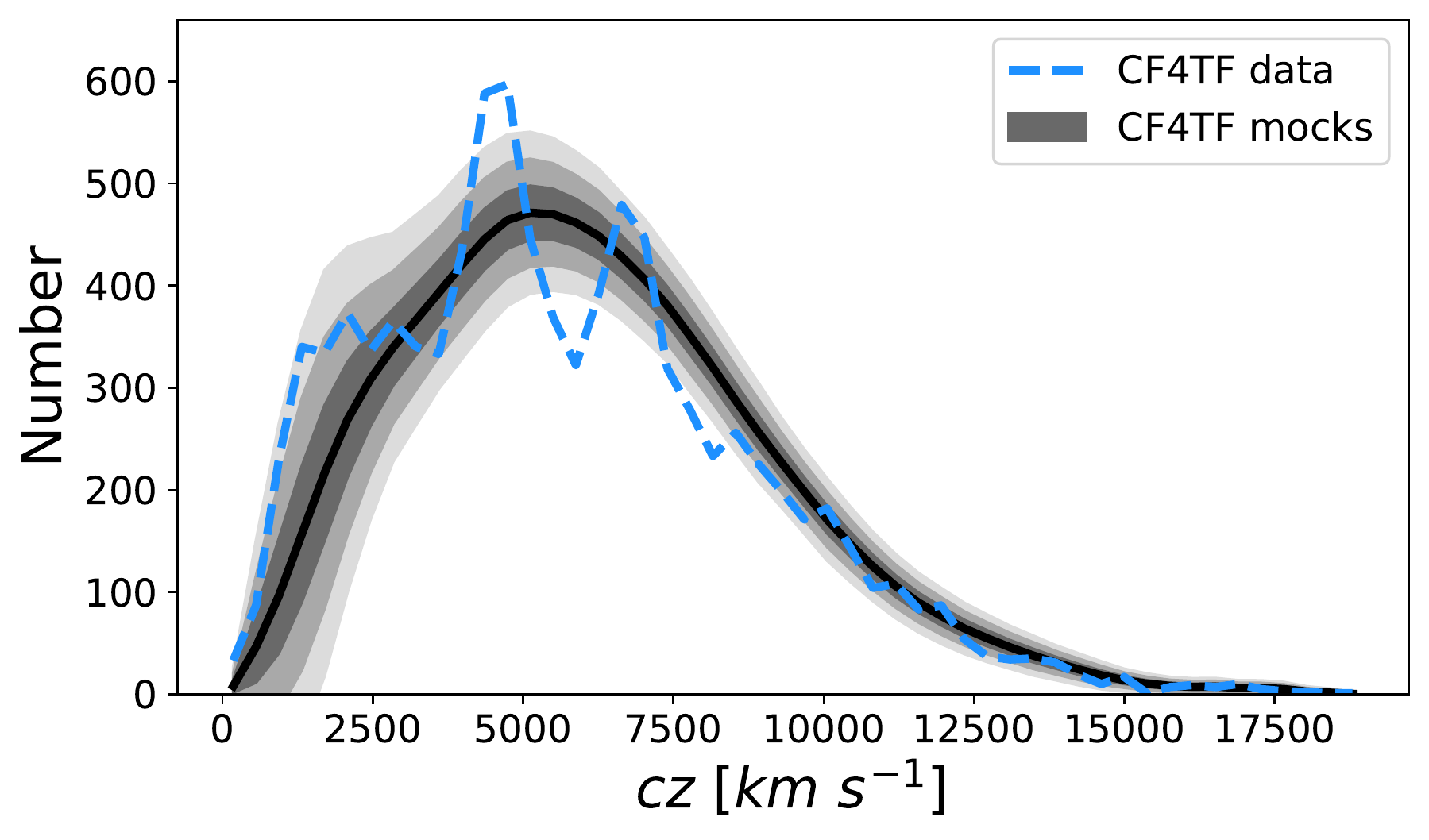}
 \caption{ Comparing the mock averages (black curves) to the real CF4TF data (blue dashed curves). The   shade areas indicate the 1$\sigma $, 2$\sigma $ and 3$\sigma $ region.   The top panels are for the distribution of i-band absolute magnitudes and log-distance ratio, respectively. The  bottom panels are for the distribution of    errors of log-distance ratios and the redshift, respectively. }
 \label{mockvsdata}
\end{figure*}

\begin{figure} 
   \includegraphics[width=\columnwidth]{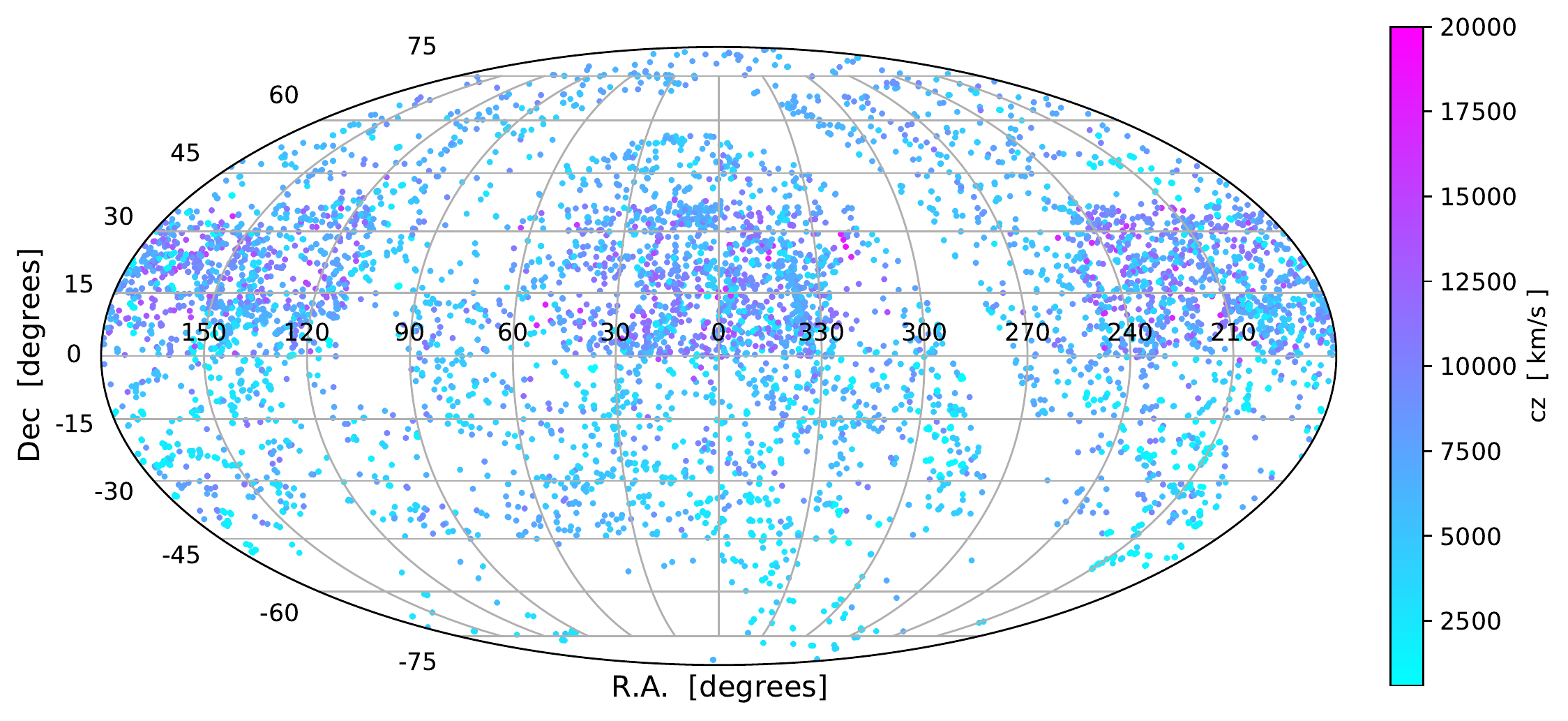}
    \includegraphics[width=\columnwidth]{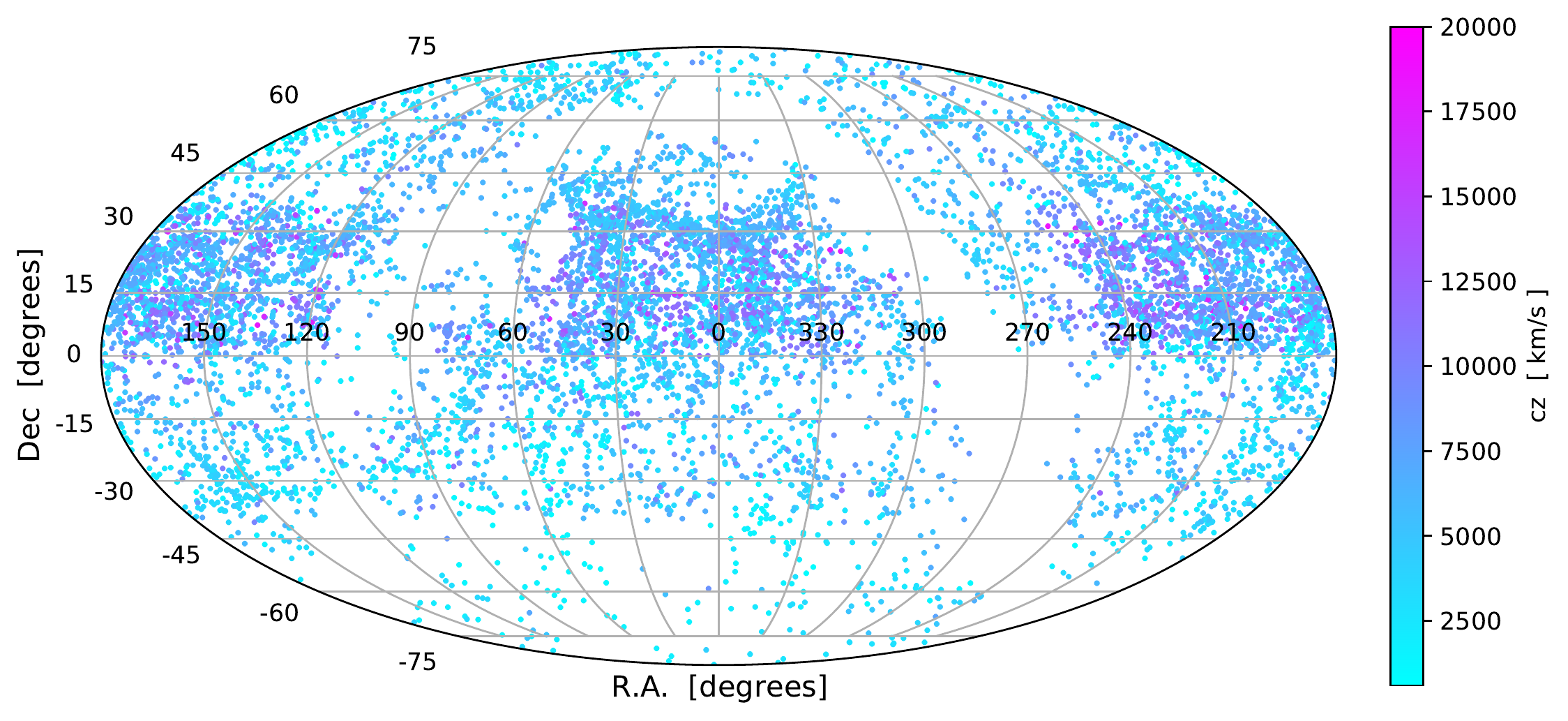}
 \caption{ The top panel shows
the sky coverage of an example mock under equatorial coordinates. The bottom  panel shows the sky coverage of the CF4TF data under equatorial coordinates. The color of the dots represent the redshift of galaxies based on the color bars.}
 \label{mockvsdatasky}
\end{figure}

Fig.\ref{mockvsdata} shows the comparison between the real CF4TF data and the mocks. 
The top left-hand-side panel shows the i-band absolute magnitude distribution. The black curve is the average of 2000 mocks, the shaded areas indicate the 1$\sigma $, 2$\sigma $ and 3$\sigma $ region. Here $\sigma$ denotes the standard deviation of the magnitudes distributions  of the 2000 mocks. The dashed blue curve is for the real CF4TF data.  The top right-hand-side panel shows the distribution of the log-distance ratio for the data and mock average. The bottom left-hand-side panel shows the distribution of the error of log-distance ratio for the data and mock average. 
The bottom  right-hand-side panel shows the redshift distribution of the data and mock average. 
In Fig.\ref{mockvsdatasky}, the top   panel shows
the sky coverage of an example mock under equatorial coordinates. For comparison, the bottom  panel shows the sky coverage of the real CF4TF data under equatorial coordinates. The color of the dots indicates the redshift based on the color bars.

The  mocks  are  in good  agreement  with  the  data. They will be continuously updated based on the 
updates of the real CF4TF catalogue and L-PICOLA simulations.

\section{Bulk and shear moments measurements} \label{sec:BKm}
 
\subsection{Bulk and shear moments} 

Using the Taylor series expansion, the line-of-sight peculiar 
velocity field, $v(d_h)$ can be expanded to the first-order
\citep{Jaffe1995,Parnovsky2001,Feldman2008,Feldman2010,Qin2019a}
\be\label{vUg}
v(d_h)=\sum^9_{p=1}U_pg_p(d_h)~,
\ee
where $U_p$ are the nine moment components
\be
U_p=\{ B_x,B_y,B_z, Q_{xx},Q_{yy},Q_{zz},Q_{xy},Q_{xz},Q_{yz} \}~,
\ee
and where the zeroth-order vector $[B_x,B_y,B_z]$ is the so called bulk flow velocity. The first-order symmetric  tensor $Q_{ij}$, $(i,j=x,y,z)$ is the so called shear moment.  
The mode functions are given by
\be\label{gps}
\begin{split}
g_p(d_h)=\{
{\bf \hat{r}}_x,  
{\bf \hat{r}}_y, 
{\bf \hat{r}}_z, 
d_h{\bf \hat{r}}_x^2, 
d_h{\bf \hat{r}}_y^2,
d_h{\bf \hat{r}}_z^2,\\
2d_h{\bf \hat{r}}_x{\bf \hat{r}}_y, 
2d_h{\bf \hat{r}}_x{\bf \hat{r}}_z, 
2d_h{\bf \hat{r}}_y{\bf \hat{r}}_z 
\}.
\end{split}
\ee
where the true comoving distance $d_h$ of a galaxy is given by
\be\label{Dz}
d_{h}(z_{h})=\frac{c}{H_0}\int_0^{z_{h}}\frac{dz'}{\sqrt{\Omega_m(1+z_h)^3+\Omega_{\Lambda}}},
\ee
and where $z_h$ is the Hubble recessional redshift of that galaxy. $H_{0}$ is the Hubble constant of present-day Universe.  
$\Omega_{m}$ and $\Omega_{\Lambda}$ are the matter and dark energy densities of present-day Universe, respectively.

\subsection{Bulk and shear moments estimator}

Measuring cosmic flows can provide us with an intuitive understanding of the amplitude and direction of the measured (not the modeled or reconstructed) velocity field. In addition, comparing the measured $U_p$ to the $\Lambda$CDM prediction enables us to test whether     $\Lambda$CDM  accurately  describe  the  motion  of galaxies in the nearby Universe.  
 To avoid the non-Gaussianity of the peculiar velocities, we use the so-called $\eta$MLE \citep{Nusser1995,Nusser2011,Qin2018,Qin2019a} to estimate the nine moment components $U_p$ from the full sky galaxy catalogue CF4TF.  

 Assuming the log-distance ratio of the $n$-th galaxy, $\eta_n$ have Gaussian errors $\epsilon_n$, then the Gaussian likelihood function of a set of $N$ galaxies can be written as \citep{Nusser1995,Nusser2011,Qin2018,Qin2019a}
\be\label{pvpi}
P(  U_p, \epsilon_{\star})=\prod^{N}_{n=1}\frac{1}{\sqrt{2 \pi \left(\epsilon_{n}^2+\epsilon_{\star,n}^2 \right)}}\exp\left({-\frac{1}{2}  \frac{    (\eta_{n}-\tilde{\eta}_{n}(U_p))^2  }{  \epsilon_{n}^2+\epsilon_{\star,n}^2 }}\right),
\ee
where the intrinsic scatter of the log-distance ratio $\epsilon_{\star}$ arises from the non-linear motions of the galaxies. We set it as a free parameter to fit. 

Following the steps of the second paragraph in Section 4.1 of  \citealt{Qin2019a},
the modeled log-distance ratio $\tilde{\eta}_{n}(U_p)$ of the $n$-th galaxy is converted 
by firstly substituting Eq.~\ref{vUg} into
the usual peculiar velocity estimator \citep{Colless2001,Hui2006,Davis2014,Scrimgeour2016,Qin2018,Qin2019a}
\be \label{travp}
v=c\left(\frac{z-z_h}{1+z_h}\right)~,
\ee
to replace $v$ to calculate $d_h(U_p)$. Then we can calculate $\tilde{\eta}_{n}(U_p)$ from $d_h(U_p)$ and the observed redshift $z$ of that galaxy.

We choose  flat priors for the 10 independent parameters.
For the three bulk flow components, we use flat priors in the interval $B_i\in[-1200,1200]$ km s$^{-1}$.
For the six shear components, we use flat priors in the interval $Q_{ij}\in[-100,100]$ h km s$^{-1}$ Mpc $^{-1}$. We use flat priors in the interval
$\epsilon_{\star}\in[-1000,1000]$ h km s$^{-1}$ Mpc $^{-1}$ for $\epsilon_{\star}$. Combining these flat priors with the likelihood in Eq.\ref{pvpi} to obtain the posterior, then we use the Metropolis-Hastings Markov chain Monte Carlo\footnote{The {\tiny PYTHON } package {  emcee } \citep{Foreman-Mackey2013} is used.} (MCMC) algorithm to estimate the parameters. 



The measurement error of the bulk flow component,  $e_{B_i}$ ($i=x,y,z$) is calculated from the standard deviation of the MCMC samples of the corresponding MCMC chain. They can be converted to the error of the bulk flow amplitude, $e_{B}$ using \citep{Scrimgeour2016,Qin2018}:
\be\label{bke2}
e^2_B=JC_{ij}J^T~,~~(i=1,2,3)~,
\ee
where $J$ is the Jacobian $\partial B/\partial B_i$. The covariance of the bulk flow components $C_{ij}$ is computed  using the MCMC samples.

\subsection{Testing on mocks}\label{sec:mocktest}


To explore how well the estimator recover the true moments, we test it using mocks (  comparison between data and cosmological models is presented in Section~\ref{sec:Result} ). 
The true moment $U_{p,t}$ of the mocks is defined as the weighted average of the true velocities of the galaxies  
 \be 
 U_{p,t}=\sum^N_{n=1} w_{p,n} {\bf v}_{n,t} \cdot {\bf \hat{r}}_n~,
 \ee
 where ${\bf v}_{n,t} $ is the true velocity of the $n$-th galaxy and is known from the simulation, ${\bf \hat{r}}_n$ is the unit vector point to that galaxy. The weights are calculated using the mode function Eq.\ref{gps}, given by 
 \be \label{aijml3}
w_{p,n}=\sum^9_{q=1}A_{pq}^{-1}\frac{g_{q,n}}{\alpha^2_n+\alpha^2_{\star}}~,~~
A_{pq}=\sum^N_{n=1}\frac{g_{p,n}g_{q,n}}{\alpha^2_n+\alpha^2_{\star}}~,
\ee 
where 
$\alpha_n$ is given by \citep{Hui2006,Johnson2014,Adams2017}:  
\be \label{vperrss}
\alpha_n=\frac{\ln(10)cz_n}{1+z_n}\epsilon_n,
\ee
which is the measurement error of peculiar velocity. 
$\alpha_{\star}$ is converted from $\epsilon_{\star}$ using the similar expression.

\begin{figure*} 
\centering
 \includegraphics[width=170mm]{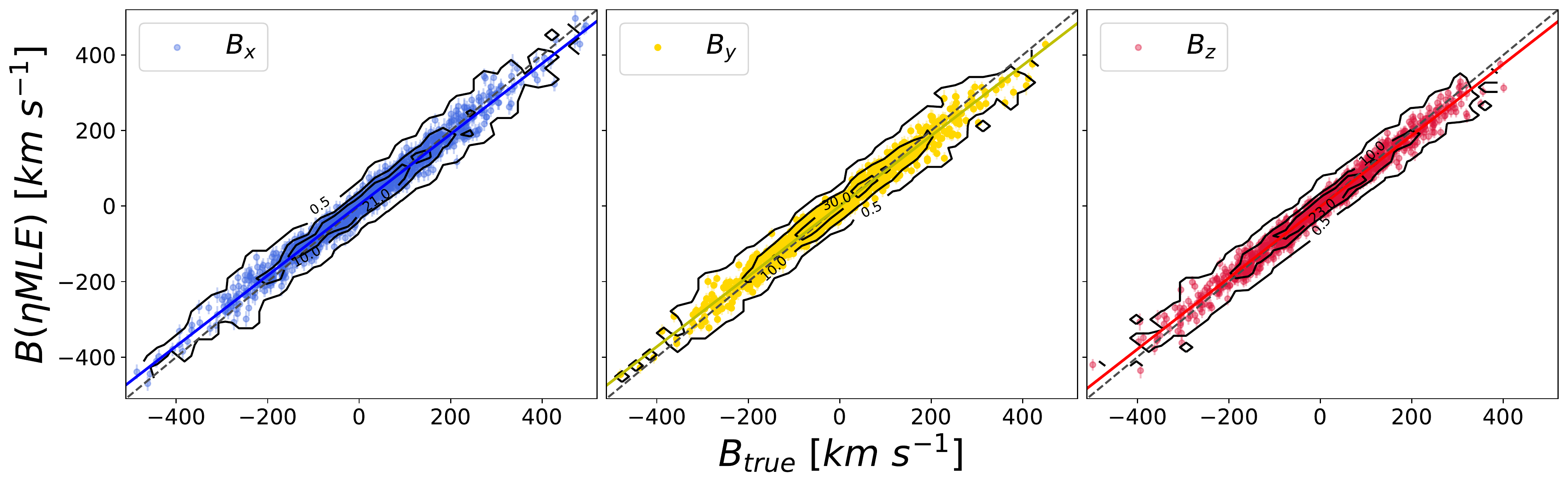}
  \includegraphics[width=170mm]{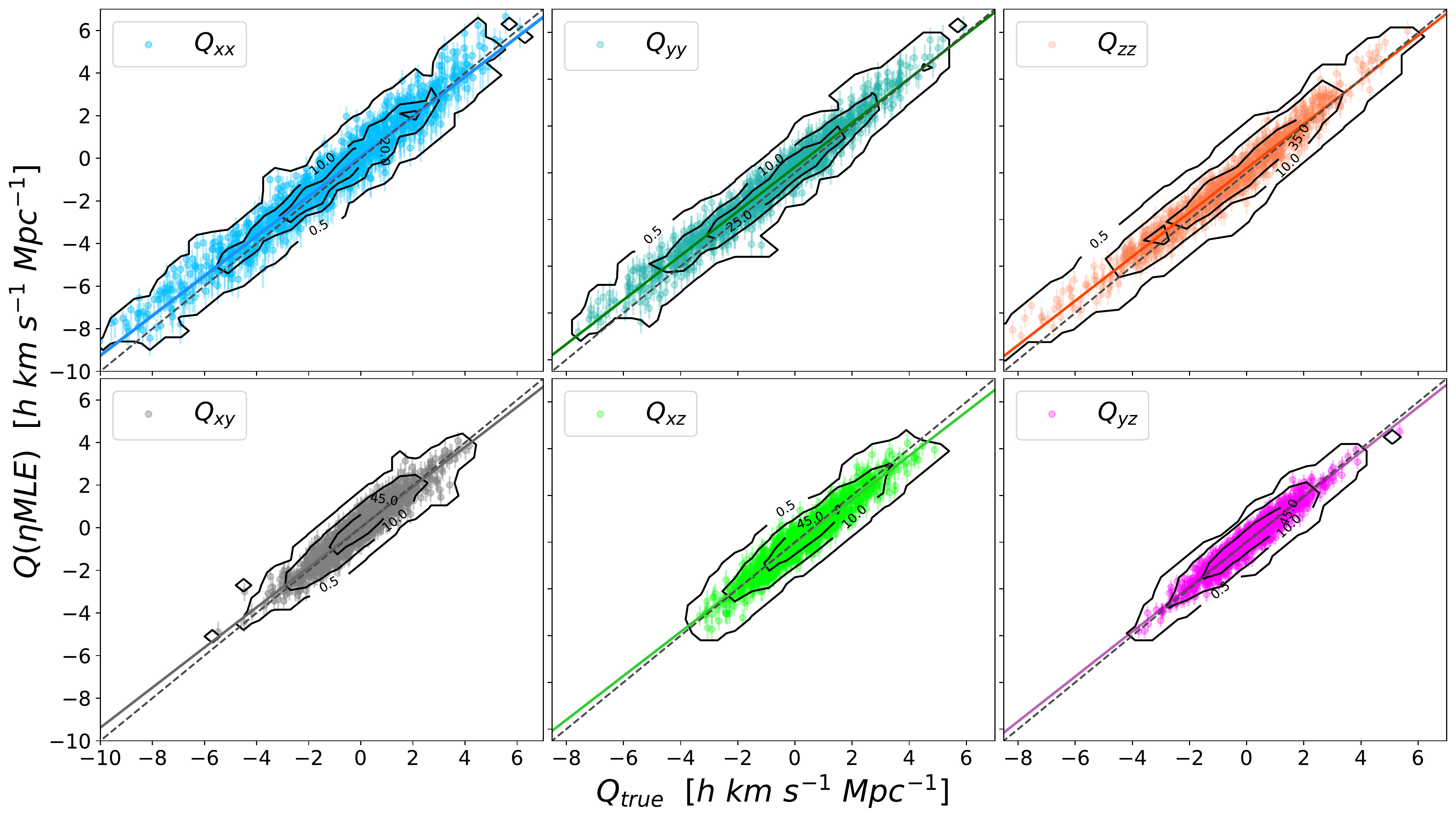}
 \caption{ Comparing the measured bulk flow velocities (top panels) and shear moments (middle and bottom panels) of CF4TF mocks to the true moments. Measured under equatorial coordinates.   1000 example mocks are shown. The black dashed lines are the  expected one-to-one relations. The colored solid lines are the best fit to the co-responding colored point. The contour indicates the 2D histogram of the dots. The numbers on the contour are the average numbers of the dots on the contour lines.  }
 \label{bmocks}
\end{figure*}

Fig.\ref{bmocks} shows the bulk (top panels) and shear moments (middle and bottom panels) measured from 
the mocks under equatorial coordinates. 
 The error bars are the measurement errors calculated using the MCMC samples, and hence include the effect of peculiar velocity measurement errors, but not cosmic variance. The impact of cosmic variance can instead be inferred from the spread in true bulk flows between mock catalogues. When comparing our results to cosmological models in Section~\ref{sec:Result}, we include both measurement errors \textit{and} cosmic variance.
 1000 example mocks are shown here. The black dashed lines are the  expected one-to-one relation. The colored solid lines are the best fit to the co-responding colored points (moments). The best fit lines are almost consistent with the one-to-one relation, indicating the $\eta$MLE can recover the true moments.  The slopes of the dashed lines are slightly different from the one-to-one relation. The reason for this is most likely due to we assume that the cosmic flow field is represented simply as bulk and shear moments, without higher order components. We leave testing of this hypothesis for future work since accurate measurements of higher moments need both larger numbers of galaxies and more isotropic and homogeneous sky coverage of galaxies.  

The
reduced $\chi^2$ between the measured moments and true moments is given by
\be\label{chi23ww}
\chi^2_{\text{red}}=\frac{1}{9\times   1000-1}(\boldsymbol{U}_m-\boldsymbol{U}_{t})\boldsymbol{\mathsf{C}}^{-1}(\boldsymbol{U}_m-\boldsymbol{U}_{t})^T ,
\ee
where the measured moments $\boldsymbol{U}_m$ and the true moments $\boldsymbol{U}_t$ 
contain 9000 elements (1000 mocks times 9 moments ), respectively.
$\mathsf{C}$ is the covariance matrix.  Eq.\ref{vUg} gives the nine moments $U_p$ as the components of the cosmic flow field. Since these moments are all being measured from a single observation location, and with incomplete sky-coverage, uncertainty of the measurement of one of these  moments will be correlated with those of another, and this is true for all moments. 
$\mathsf{C}$ is a 9000$\times$9000 matrix with 1000 9$\times$9 
diagonal blocks and zero elsewhere. The 1000 diagonal blocks are calculated from the MCMC chains.
The $\chi^2_{red}=1.806$. 
This slightly larger value of $\chi^2_{red}$ 
is due to the intrinsic scatter between the measured moments and true moments, which is arise from the intrinsic scatter of the peculiar velocities $\alpha_{\star}$ (or $\epsilon_{ \star}$ of log-distance ratio). $\alpha_{\star}$ (or $\epsilon_{ \star}$) is to encapsulate the non-linear peculiar motions of galaxies. The peculiar velocity estimator  
Eq.\ref{travp} is not good enough to account for the non-linear peculiar motions. Therefore a more robust
peculiar velocity estimator need to be developed in future work.


\section{Results and Discussion} \label{sec:Result}

\subsection{Results and comparing to $\Lambda$CDM prediction}
\begin{figure*} 
\centering
  \includegraphics[width=120mm]{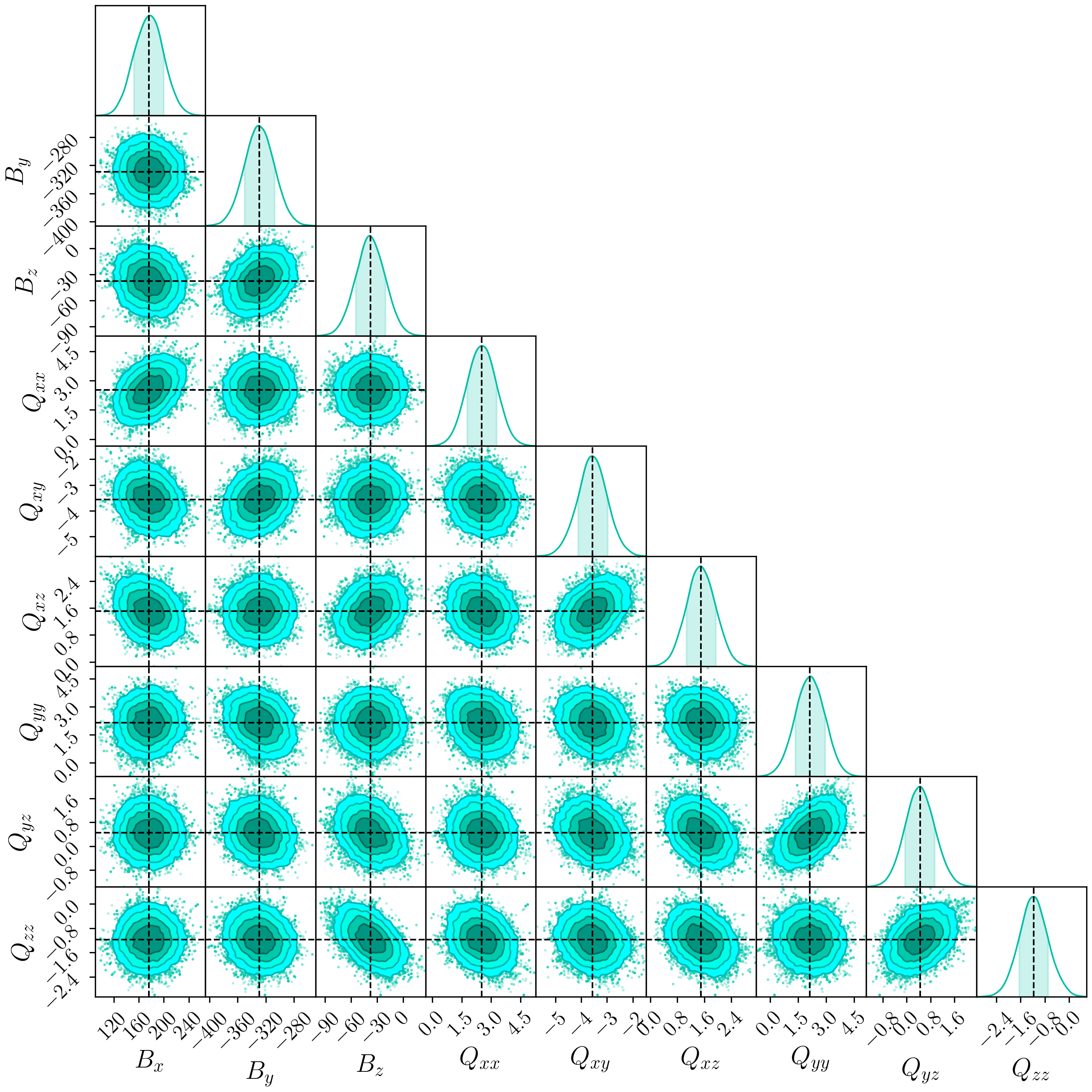}
 \caption{ The 2D contours  and  the marginalised histograms of the MCMC samples of the bulk and shear moments. The shade areas in the histograms indicates the 68\% confidence level of the moments.}
 \label{mcbq}
\end{figure*}

Applying $\eta$MLE to the real CF4TF data, we obtain the measured bulk and shear moments in the local Universe, as presented in Table \ref{bkflb}. The moments are measured under Galactic coordinates. Fig.\ref{mcbq} shows the 
2D contours and histograms of the 
MCMC samples of the moments. The shaded areas of the histograms indicate the 1$\sigma$ errors of the moments.  
To compare to theory, the characteristic scale of cosmic flow measurement  is defined as \citep{Scrimgeour2016}
\be\label{depthri}
d_{\text{MLE}}=\frac{\sum |\boldsymbol{d}_{h,n}|W_n}{\sum W_n}~,
\ee
where the weight factors $W_n=1/(\alpha_n^2+\alpha^2_{\star})$.

The $\Lambda$CDM predicted moments should have
zero mean and `cosmic root mean square' (CRMS) variation \citep{Feldman2010}. The CRMS is calculated from the following equation
\citep{Feldman2010,Ma2011,Johnson2014}
\be \label{Rcrms}
R_{pq}^{v}=\frac{\Omega^{1.1}_mH^2_0}{2\pi^2} \int\mathcal{W}_{pq}^2(k)\mathcal{P}(k)dk~,
\ee
where the matter density power spectrum $\mathcal{P}(k)$ is generated using the \textsc{CAMB} pakage \citep{Lewis:1999bs}. The window function is given by
\citep{Feldman2010,Ma2011,Johnson2014}:
\be\label{wdfcs}
\mathcal{W}_{pq}^2(k)=\sum_{m,n}^Nw_{p,m}w_{q,n}f_{mn}(k) ~.
\ee
where $f_{mn}(k)$ is given by \citep{Ma2011,Johnson2014}
\be 
\begin{split}
f_{mn}(k)=&\frac{1}{3}\left[ j_0(kD_{mn}) -2j_2(kD_{mn})  \right] \hat{\bf r}_m \cdot \hat{\bf r}_n \\
+& \frac{1}{D^2_{mn}}j_2(kD_{mn})r_mr_n\sin^2(\alpha_{mn}),
\end{split}
\ee 
and where $D_{mn}\equiv|{\bf r}_m-{\bf r}_n|$ and $\alpha_{mn}=\cos^{-1}(\hat{\bf r}_m \cdot \hat{\bf r}_n)$. ${\bf r}_n$ is the position vector of the $n$-th galaxy. $j_0(x)$ and $j_2(x)$ are the 0th- and 2nd-order  spherical Bessel function of the first kind. 
The weight factors are given by Eq.\ref{aijml3}.

\begin{figure} 
 \includegraphics[width=\columnwidth]{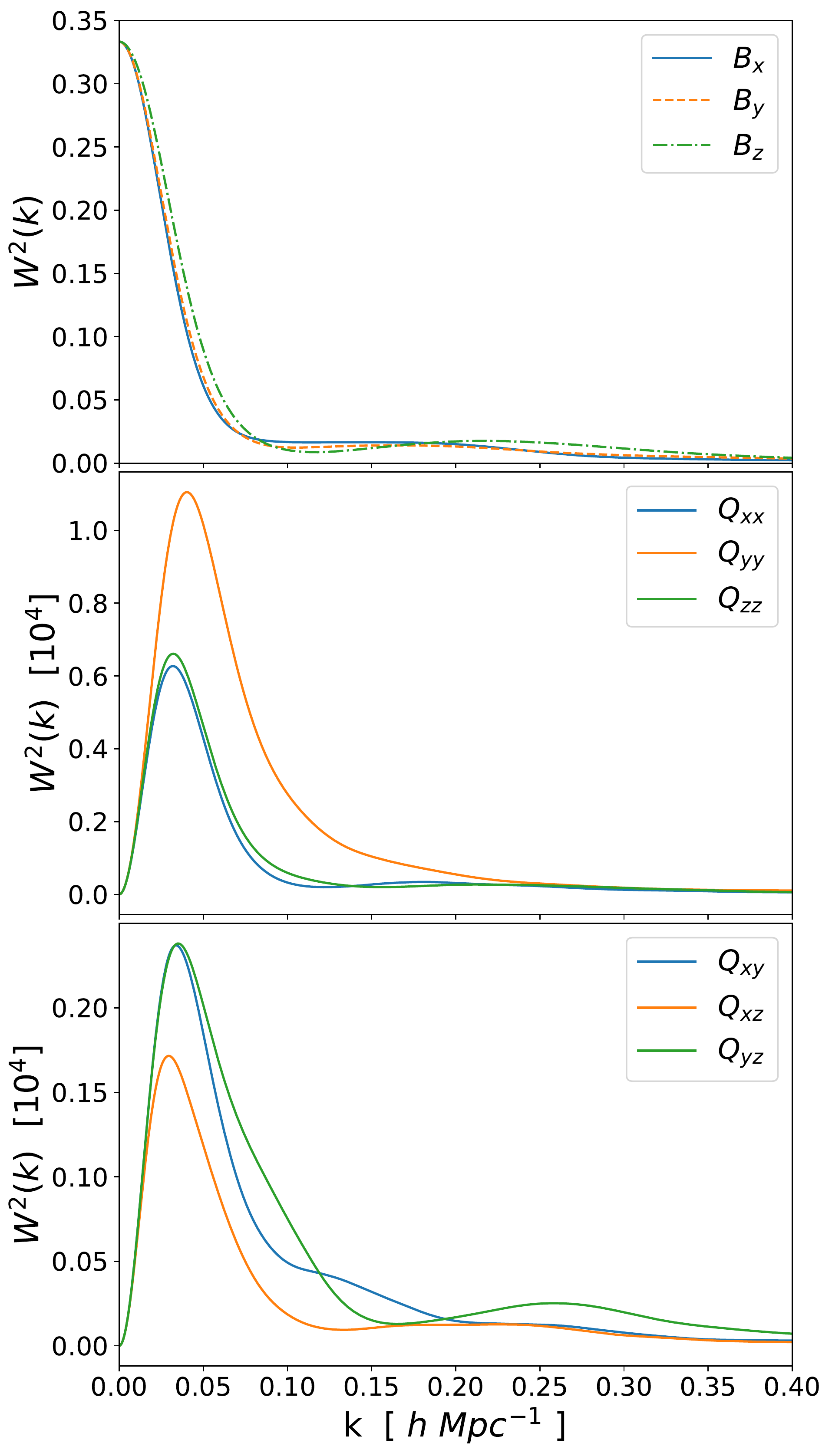}
 \caption{The window functions for CF4TF under Galactic coordinates. The top panels are for the bulk flow components. The middle panels are for the diagonal elements of the shear moments . The bottom panels are for the off-diagonal elements of the shear moments.}
 \label{WinFij}
\end{figure}

 Fig.\ref{WinFij} shows the $p=q$ components of the window function $W^2_{pq}$ of the CF4TF data under Galactic coordinates. 
 In Table~\ref{bkflb}, we list the CRMS predicted by  $\Lambda$CDM.
As shown in the top panel of Fig.\ref{WinFij}, the amplitudes of window functions for the three bulk flow components are similar, therefore, the co-responding CRMS of the three bulk flow components are similar in Table ~\ref{bkflb}. As shown in the middle and bottom panels of  Fig.\ref{WinFij}, the window functions for the $y$ direction related components [$Q_{yy}$, $Q_{xy}$, $Q_{yz}$] have larger amplitudes, indicating larger cosmic variance in this direction. This is due to the an-isotropic sky coverage caused by the denser region contributed by ALFALFA galaxies. Therefore, in Table ~\ref{bkflb}, the CRMS for the $y$ direction related components, especially $Q_{yy}$, are larger then other shear moments components.

The $\Lambda$CDM predicted moments should have
zero mean. Therefore, the $\chi^2$ between the 9 measured moments and $\Lambda$CDM prediction is given by:
\be 
\chi^2=U_p(C_{pq}+R^v_{pq})^{-1}U^T_q~.
\ee 
where $C_{pq}$ is the covariance of measurement errors, it is calculated from the MCMC samples. The $\chi^2=10.4$, and the corresponding $p$-value is 0.319, indicating the measurements are consistent with the $\Lambda$CDM prediction. 

\begin{table}   \small
\caption{ The bulk and shear moments measured from the CF4TF data. To compare to $\Lambda$CDM prediction, we also list the CRMS in the last column.  The number of degrees of freedom is 9.}
\begin{tabular}{|c|c|c|}
\hline
\hline

     & $\eta$MLE & CRMS  \\
\hline
  $B_x$ (km s$^{-1}$)  & $175.9\pm23.5$ & $\pm$173.4  \\
\hline
  $B_y$ (km s$^{-1}$)  & $-330.2\pm21.9$ & $\pm$175.1  \\
\hline
  $B_z$ (km s$^{-1}$)  & $-38.9\pm17.1$ & $\pm$183.5  \\
\hline
  $Q_{xx} $  ($h$ km s$^{-1}$ Mpc$^{-1}$)   & $2.53\pm0.74$ & $\pm$2.62\\
\hline
  $Q_{xy} $  ($h$ km s$^{-1}$ Mpc$^{-1}$)  & $-3.58\pm0.58$ & $\pm$1.70\\
\hline
  $Q_{xz} $  ($h$ km s$^{-1}$ Mpc$^{-1}$)   & $1.52\pm0.43$ & $\pm$1.43  \\
\hline
  $Q_{yy} $  ($h$ km s$^{-1}$ Mpc$^{-1}$)  & $2.14\pm0.79$ & $\pm$3.65  \\
\hline
  $Q_{yz}$($h$ km s$^{-1}$ Mpc$^{-1}$)   & $0.47\pm0.51$ & $\pm$1.77  \\
 \hline
  $Q_{zz} $  ($h$ km s$^{-1}$ Mpc$^{-1}$)   & $-1.16\pm0.48$ & $\pm$2.72   \\
 \hline
 $\chi^2$ & \multicolumn{2}{|c|}{10.4}\\
  \hline
  $p$-value &	\multicolumn{2}{|c|}{0.319}\\
  	\hline
 Direction of bulk flow & \multicolumn{2}{|c|}{$l$=298.4$\pm$3.4$^{\circ}$, $~b$= -5.9$\pm$2.7$^{\circ}$ }  \\  
      \hline
$d_{MLE}$   ($h^{-1}$ Mpc) & \multicolumn{2}{|c|}{35}  \\  
      \hline
   
\end{tabular}
 \label{bkflb}
\end{table}


\subsection{Comparing to other literature}

To visualise the comparison of different bulk flow measurements from other surveys on  a  single  figure,  we need to  standardise the window function due to the differing geometries and depths of the surveys. In this work, we use the spherical top-hat window function: 
\be 
\mathcal{W}(k)=3(\sin kR-kR\cos kR)/(kR)^3. 
\ee
The other surveys we will compare are:
W09: \citet{Watkins2009};
C11: \citet{Colin2011};
D11: \citet{Dai2011}; 
N11: \citet{Nusser2011};
T12: \citet{Turnbull2012}; 
M13: \citet{Ma2013};
H14: \citet{Hong2014};  
S16: \citet{Scrimgeour2016};  
Q18: \citet{Qin2018}; 
Q19: \citet{Qin2019a};
B20: \citet{Boruah2020};
S21: \citet{Stahl2021}.

\begin{figure*} 
\centering
 \includegraphics[width=175mm]{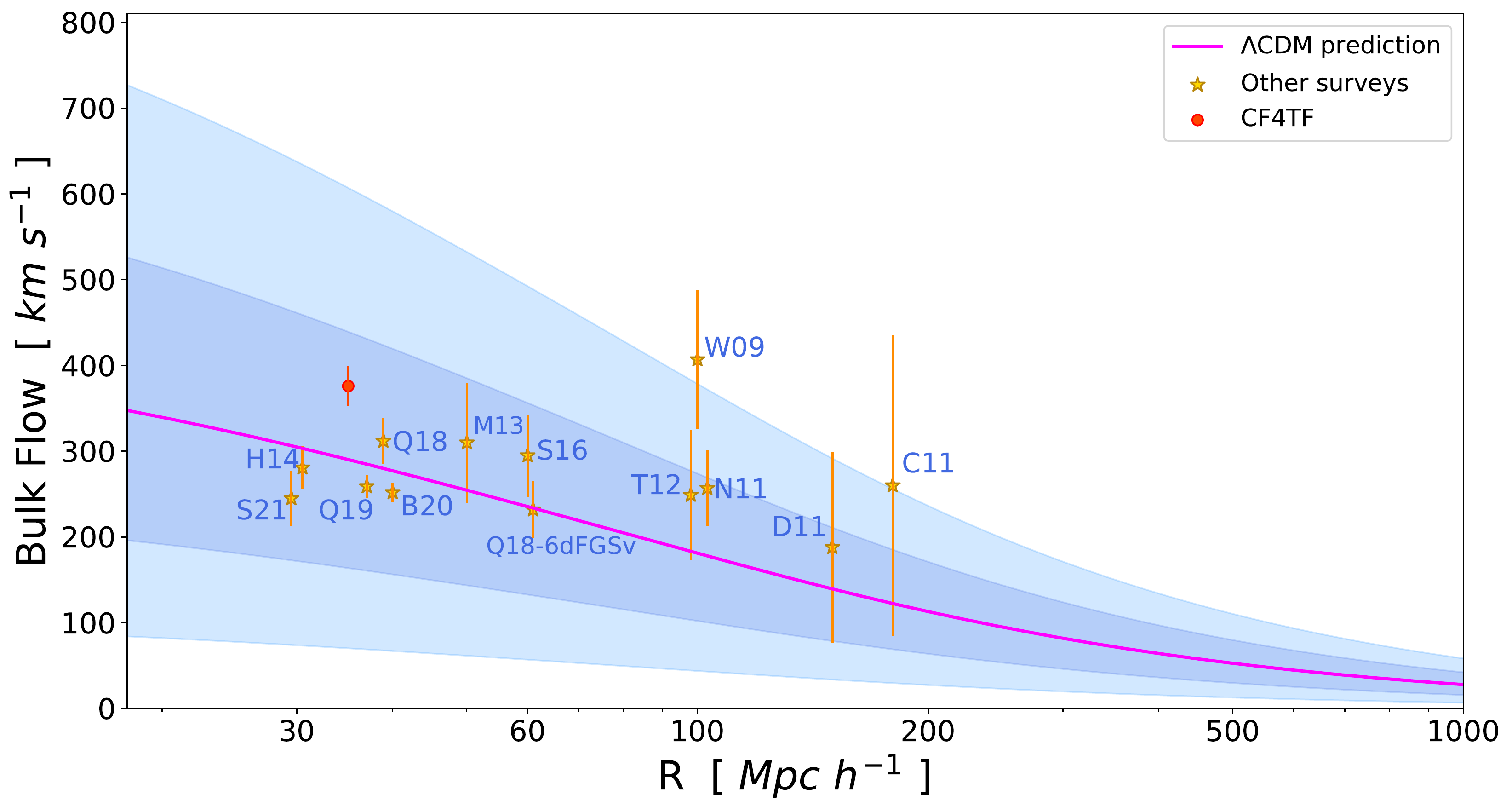}
 \caption{  Comparing the bulk flow measurements from different surveys:W09: \citet{Watkins2009};
  C11: \citet{Colin2011};
D11: \citet{Dai2011}; 
N11: \citet{Nusser2011};
T12: \citet{Turnbull2012}; 
M13: \citet{Ma2013};
H14: \citet{Hong2014};  
S16: \citet{Scrimgeour2016};  
Q18: \citet{Qin2018}; 
Q19: \citet{Qin2019a};
B20: \citet{Boruah2020};
S21: \citet{Stahl2021}.  The pink curve is the $\Lambda$CDM prediction calculated from the spherical top-hat window function. The shade areas indicates the 1$\sigma$ and 2$\sigma$ cosmic variance. The yellow stars are the measurements from other surveys. 
The red dot is our measurement using CF4TF. }
 \label{bvxRR}
\end{figure*}

In Fig.\ref{bvxRR}, the pink curve is the $\Lambda$CDM prediction calculated from the spherical top-hat window function. The shade areas indicates the 1$\sigma$ and 2$\sigma$ cosmic variance. The yellow stars are the measurements from other surveys. 
The red dot is our measurement using CF4TF. 
Following the arguments in \citealt{Scrimgeour2016}, 
to be comparable to the top-hat window function prediction,
we plot the W09 and T12 measurements at twice their quoted depth since  they have 
Gaussian windows. 
S16 measures the bulk flow from 
the 6dFGSv data \citep{Springob2014}, which is a hemispherical top-hat.  Following the arguments in S16, we plot their measurement  
at effective radius $R_{\text{eff}}=(R^3/2)^{1/3}$. Q18 revised the bulk flow measurement of 6dFGSv by calibrating a bias in the Malmquist bias correction of 6dFGSv distance measurements, the revised result is show in Q18-6dFGSv. Generally, most of the measurements are consistent with the $\Lambda$CDM prediction, while the W09 does not agree with  the $\Lambda$CDM prediction. 

Comparing to other measurements (yellow stars), our measurement (red dot) has smaller error. B20 has smallest error bar, they measure bulk flow by comparing the reconstructed velocity field of 2M++ \citep{Carrick2015} to the measured velocity field of 465 supernovas. 
However, as presented in Section 5.1 of \citealt{Carrick2015}, the reconstructed field has many potential systematic effects which are hard to test. 
Q19 measures the bulk flow using Cosmicflows-3 catalogue \citep{Tully2016} which contains 391 Type Ia supernovas, therefore, has smaller error bar too. However, the major part of the published Cosmicflows-3 catalogue is the 6dFGSv, which has not perfect Malmquist bias correction \citep{Qin2018}, therefore this measurement has  potential systematic errors.

\begin{figure*} 
\centering
 \includegraphics[width=175mm]{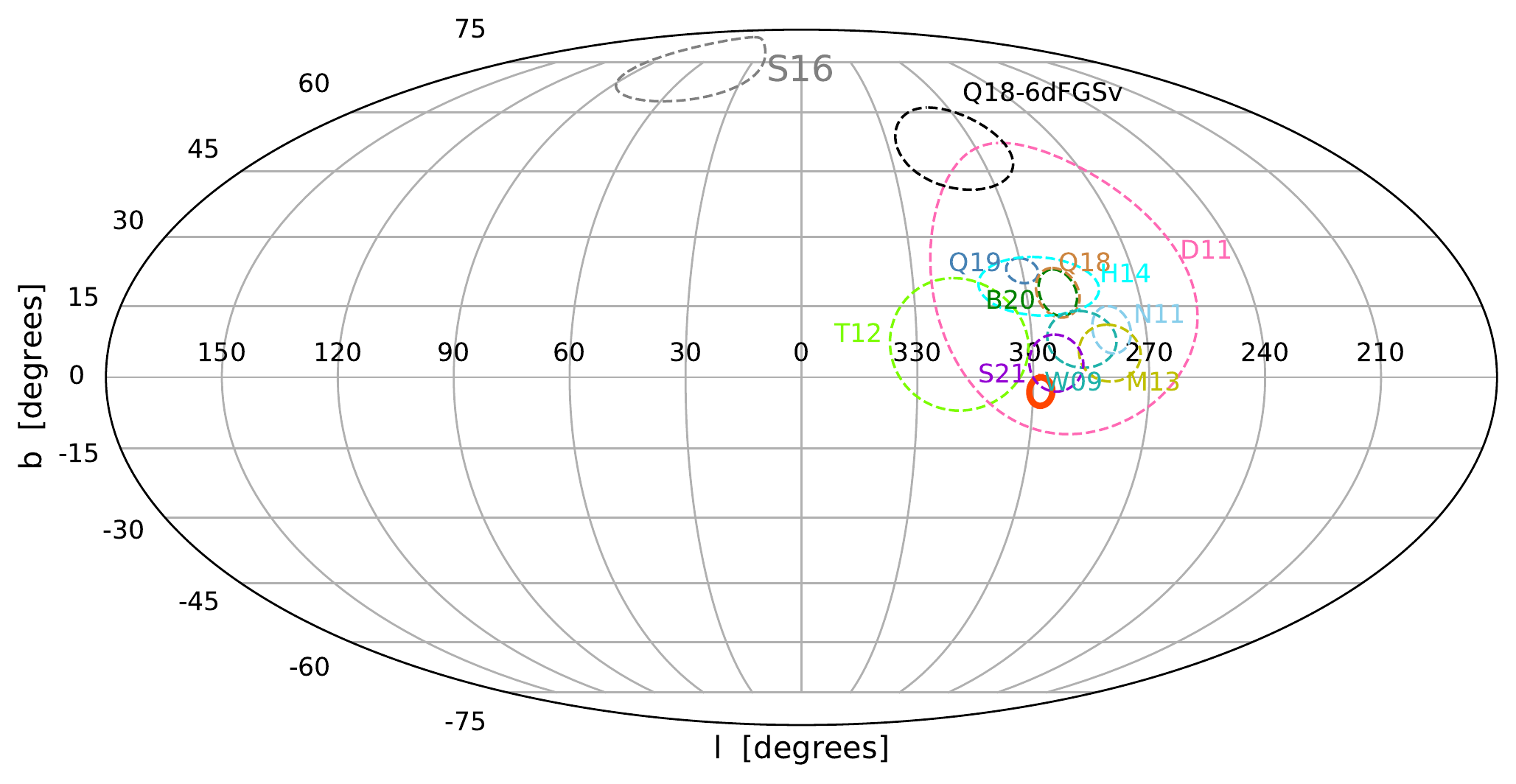}
 \caption{ Comparing the bulk flow direction from different surveys under Galactic coordinates:W09: \citet{Watkins2009};
D11: \citet{Dai2011}; 
N11: \citet{Nusser2011};
T12: \citet{Turnbull2012}; 
M13: \citet{Ma2013};
H14: \citet{Hong2014};  
S16: \citet{Scrimgeour2016};  
Q18: \citet{Qin2018}; 
Q19: \citet{Qin2019a};
B20: \citet{Boruah2020};
S21: \citet{Stahl2021}.  The red solid circle  is   our measurement using CF4TF.  
The size of the circles indicate the 1$\sigma$ error of the measurements. }
 \label{lbcompr}
\end{figure*}

Fig.\ref{lbcompr}  shows directions of bulk flows measured from different surveys under Galactic coordinates.  The S16 measurements is significantly biased from other surveys. The revised result is show in Q18-6dFGSv. 

\section{Conclusion} \label{sec:conc}

In this paper, we present the mock sampling algorithm of CF4TF data. These mocks can realize the luminosity selection, survey geometry and clustering of the real CF4TF data. A combination of these CF4TF mocks and 6dFGSv mocks \citep{Qin2018,Qin2019b} provide us with 
the mock catalogues for the two largest subsets of the final full Cosmicflows-4 catalogue. 
The mocks
can be used to further study cosmology in  future work. For example,
estimating the covariance matrix and errors of
power spectrum and two-point correlation of velocities.
The mock catalogues underlying this article will be shared on a reasonable request to the corresponding author.

We use mocks to test the cosmic flow estimator $\eta$MLE,  we find that the estimator works well to recovers the true moments. 
We measure the bulk and shear moments of the local Universe using the CF4TF data. 
We find that the bulk flow in the local Universe is 376 $\pm$ 23 (error) $\pm$ 183 (cosmic variance) km s$^{-1}$ at depth $35$ Mpc $h^{-1}$, to the Galactic direction of $(l,b)=(298\pm 3^{\circ},  -6\pm 3^{\circ})$. Both the measured bulk and shear moments are consistent with the $\Lambda$ cold dark matter  model prediction.

\acknowledgments

FQ and DP are supported by the project \begin{CJK}{UTF8}{mj}우주거대구조를 이용한 암흑우주 연구\end{CJK} (``Understanding Dark Universe Using Large Scale Structure of the Universe''), funded by the Ministry of Science. CH and KS are supported by the Australian Government through the Australian Research Council’s Laureate Fellowship funding scheme (project FL180100168). 

%

\vspace{5mm}
\facilities{The \textsc{L-PICOLA} simulation and CF4TF mock sampling algorithm were performed on the OzSTAR national facility at Swinburne University of Technology. The Cosmicflows-4 Tully-Fisher catalogue is downloaded from the Extragalactic Distance Database (EDD) \url{http://edd.ifa.hawaii.edu/}. }


\software{\textsc{emcee} \citep{Foreman-Mackey2013},  
          \textsc{ChainConsumer} \citep{ChainConsumer}, 
          \textsc{SCIPY} \citep{Virtanen2020},
          \textsc{MATPLOTLIB} \citep{Hunter2007},
          \textsc{HEALPix} \citep{healpy1,healpy2}.  }



\appendix

\section{i-band Luminosity function}\label{sec:lumi}

In the CF4TF catalogue, we have the fully corrected \footnote{The corrections are done for Milky  Way  obscuration, redshiftk-correction, andaperture effects and global dust obscuration \citep{Kourkchi2020}.} i, g, r, z, $w_1$ and $w_2$ bands apparent magnitudes. However, 
non of these bands are for all galaxies. To fit the luminosity function of all CF4TF galaxies, we need to work out a single band of magnitudes for all galaxies.

There are 6776 galaxies have i-band apparent magnitudes $m_i$, while 5037 galaxies have $w_1$-band apparent magnitudes $m_{w_1}$. 
 There are 2033 galaxies have both $w_1$-band and i-band apparent magnitudes in the CF4TF catalogue. Following the arguments in \citealt{Kourkchi2020}, we can firstly calculate the differences between the magnitudes of the two bands for these common galaxies, 
 $
 \Delta_{i,w_1}=m_i-m_{w_1}.
$ 
Then using the equation shown in Figure 15 of \citealt{Kourkchi2019} to 
convert the $w_2$-band main principal components $P_{1,w_2}$ to $w_1$-band main principal components $P_{1,w_1}$. Finally, following the arguments in Appendix B4 of \citealt{Kourkchi2020}, we using the  Random Forest Regressor from the Python
package \textsc{scikit-learn} to obtain the   $w_1$-band apparent magnitudes of all CF4TF galaxies.

\begin{figure} 
\centering
 \includegraphics[width=\columnwidth]{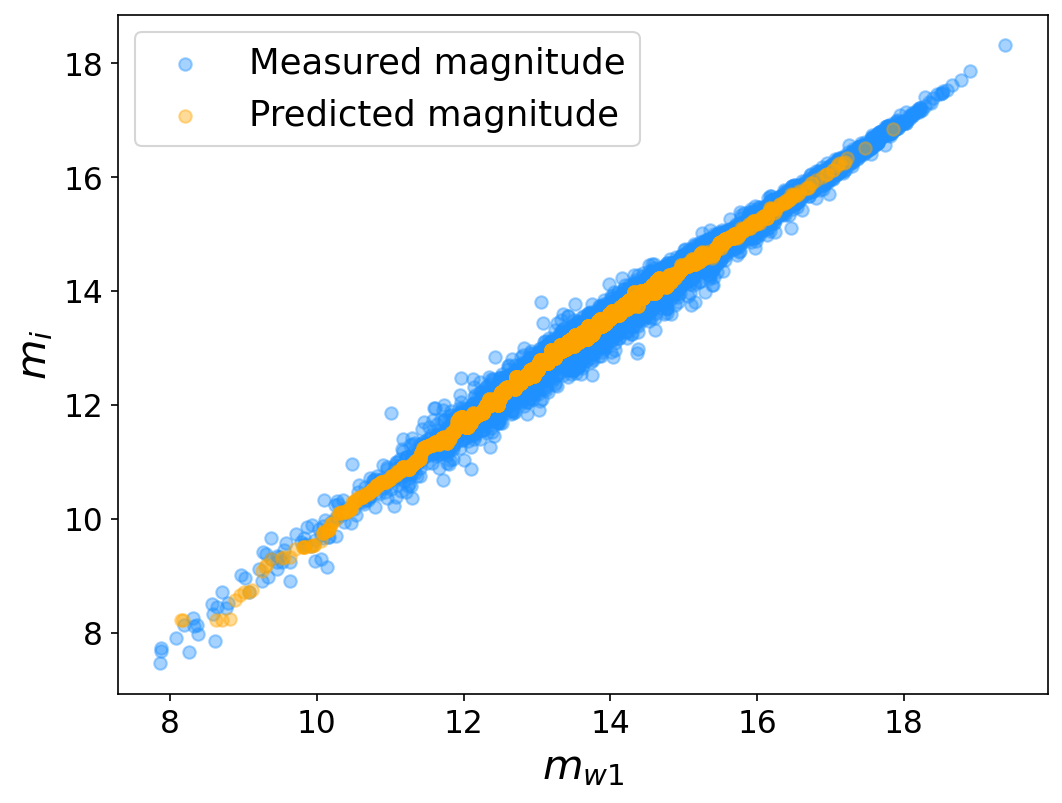}
 \caption{The relation between the i-band apparent magnitudes and $w_1$-band apparent magnitudes of CF4TF.}
 \label{lbde}
\end{figure}

\begin{figure} 
\centering
 \includegraphics[width=\columnwidth]{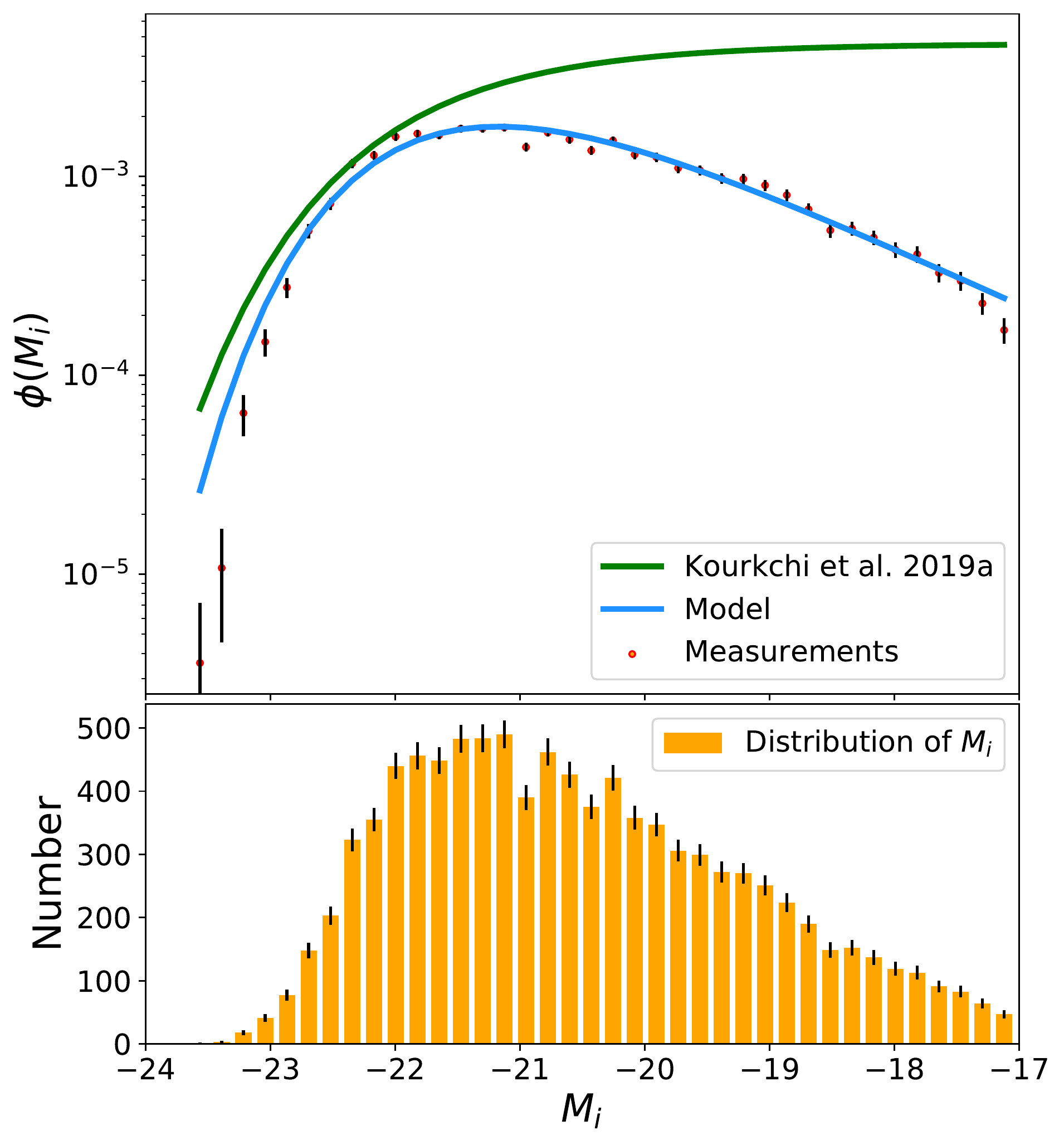}
 \caption{The bottom panel   shows the histogram of the i-band absolute magnitude. The top panel  are the corresponding luminosity function. The blue curve is the Schechter model fit to the data. The green curve is the Schechter model given by \citealt{Kourkchi2020a}.}
 \label{asfd}
\end{figure}

Since the i-band magnitude cut is the commonly quoted one in the previous  literature \citep{Kourkchi2019,Kourkchi2020,Kourkchi2020a}, we then need to convert $w_1$-band magnitudes to i-band magnitudes. 
The blue dots in Fig.\ref{lbde} shows the relation between the i-band apparent magnitudes and $w_1$-band apparent magnitudes of the 6776 galaxies. Then using Random Forest Regressor to obtain the  i-band apparent magnitudes of the rest galaxies, as shown in the yellow dots in Fig.\ref{lbde}. Thus, we obtain the i-band apparent magnitudes for all CF4TF galaxies. Converting the i-band apparent magnitudes to absolute magnitudes using the distance modulus data in the CF4TF catalogue. The distribution of i-band absolute magnitudes $M_i$ is shown in the bottom panel of Fig.\ref{asfd}.

To generate mock catalogues for CF4TF, we need to fit the i-band
luminosity function for the data. We use the  Schechter function \citep{Schechter1976}
\be 
\phi(M_i)=0.4\ln(10)\phi_{\star}10^{-0.4(M_i-M_{\star})(\alpha+1)}\exp{\left[ -10^{-0.4(M_i-M_{\star})}\right]}.
\ee 
The best fit values of parameters $M_{\star}$ and $\alpha$ are give in the 
Table 6 of \citealt{Kourkchi2020a}. However, the shape of the luminosity function given by their parameters can not match the measurements, see the green curve in the top panel of Fig.\ref{asfd}. In addition, \citealt{Kourkchi2020a} does not provide the normalization parameter $\phi_{\star}$ of the luminosity function. Therefore, 
instead of using the values in Table 6 of \citealt{Kourkchi2020a},
we fit our own luminosity function.  
The best fitted values are $M_{\star}=-21.528^{+0.030}_{-0.029}$ and $\alpha=-0.280^{+0.020}_{-0.018}$. The fit result is  shown in the blue curve in the top panel of Fig.\ref{asfd}.

To emphasize, in this stage we only want to know the shape parameters of the luminosity function, i.e. we only want to know the values of $M_{\star}$ and $\alpha$. While the normalization parameter $\phi_{\star}$ will be fitted in Section \ref{sec:mock} since it is difficult to obtain the survey volume of CF4TF due to the complex survey geometry.

\section{The five CF4TF patches}\label{sec:patch}

\begin{figure} 
\centering
 \includegraphics[width=\columnwidth]{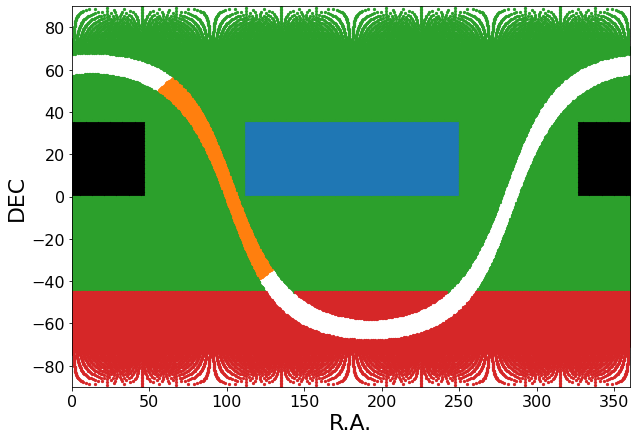}
 \caption{  The five non-overlapping patches of the CF4TF sky coverage we adopt for producing our mock catalogues.} 
 \label{asfddd}
\end{figure}

  In building our mock catalogues, we treat the inhomogeneous CF4TF data as consisting of 5 distinct patches, each of which has their own unique redshift distribution. Fig.\ref{asfddd} shows the five patches of the CF4TF sky coverage we use, which are defined as, and correspond to:
\begin{enumerate}
\item{
The blue region: 0$^{\circ}$<Dec<38$^{\circ}$ and 111$^{\circ}$<R.A.<250$^{\circ}$. Covers the ALFALFA data in the northern Galactic region.}
\item{
The black region: 0$^{\circ}$<Dec<38$^{\circ}$ and 0$^{\circ}$<R.A.<48$^{\circ}$ and 325.5$^{\circ}$<R.A.<360$^{\circ}$. Covers the ALFALFA data in the southern Galactic region}
\item{
The green region: Dec>-45$^{\circ}$. Relatively uniform galaxies, mainly from ADHI, with additions from the Springob/Cornell and Pre-Digital HI catalogues.}
\item{
The orange region: 149$^{\circ}$<$l$<255$^{\circ}$ and -5$^{\circ}$<$b$<5$^{\circ}$. Sparse data in the Galactic plane from ADHI, the Springob/Cornell and Pre-Digital HI catalogues, where Galactic extinction reduces the number density of observed TF objects.}
\item{
The red region: Dec<-45$^{\circ}$. Mainly galaxies from the ADHI but observed using the Parkes Telescope, which results in a lower number density of these galaxies relative to the green region. }
\end{enumerate} 

\bibliography{FQinRef}{}
\bibliographystyle{aasjournal}



\end{document}